\documentclass{decar-wsd}    
\usepackage{decar-common}    
\usepackage{decar-dynamics}  

\usepackage[authoryear,longnamesfirst]{natbib}

\DeclareMathAlphabet{\mbf}{OT1}{ptm}{b}{n}

\usepackage{tikz}
\usetikzlibrary{positioning,calc,decorations.pathreplacing,patterns,decorations.pathmorphing,decorations.markings}
\usepackage[percent]{overpic}
\usepackage{siunitx}
\usepackage{booktabs}

\definecolor{oiorange}{rgb}{0.90, 0.60, 0.00}
\definecolor{oiskyblue}{rgb}{0.35, 0.70, 0.90}
\definecolor{oibluishgreen}{rgb}{0.00, 0.60, 0.50}
\definecolor{oiyellow}{rgb}{0.95, 0.90, 0.25}
\definecolor{oiblue}{rgb}{0.00, 0.45, 0.70}
\definecolor{oivermillion}{rgb}{0.80, 0.40, 0.00}
\definecolor{oireddishpurple}{rgb}{0.80, 0.60, 0.70}

\def\tsc#1{\csdef{#1}{\textsc{\lowercase{#1}}\xspace}}
\tsc{WGM}
\tsc{QE}

\usepackage{amsmath}
\usepackage{graphicx}
\usepackage{float}
\usepackage{caption}
\usepackage{lipsum}
\usepackage[hidelinks]{hyperref}

\title{Forward-Backward Extended DMD with an Asymptotic Stability Constraint}
\begin{document}

\makeatletter
\def\@maketitle{%
  \null
  \vskip 2em%
  \begin{center}%
  \let \footnote \thanks
    {\LARGE \textbf{\@title} \par}%
    \vskip 1.5em%
    {\large
      \lineskip .5em%
      \begin{tabular}[t]{c}%
        \emph{Louis Lortie\thanks{Group for Research in Decision Analysis (GERAD), Montreal QC H3T 1N8, Canada; McGill University, Montreal QC H3A 0C3, Canada; Department of Mechanical Engineering, McGill University, Montreal QC H3A 0C3, Canada (\url{louis.lortie@mail.mcgill.ca}, \url{steven.dahdah@mail.mcgill.ca},
\url{james.richard.forbes@mcgill.ca)}}, Steven Dahdah\footnotemark[1], and James Richard Forbes\footnotemark[1]} \\
        \small{Department of Mechanical Engineering, McGill University} \\
        \small{817 Sherbrooke Street West, Montreal QC H3A 0C3} \\
      \end{tabular}\par}%
    \vskip 1em%
  \end{center}%
  \par
  \begin{center}%
    \rule{\linewidth}{1.5pt}%
  \end{center}%
  \raggedright}
\makeatother

\maketitle


\begin{abstract}
This paper presents a data-driven method to identify an asymptotically stable Koopman system from noisy data. In particular, the proposed approach combines system inputs and approximations of the system's forward- and backward-in-time dynamics to reduce bias caused by noisy data while enforcing asymptotic stability. A Koopman model of an inherently asymptotically stable system can be unstable due to noisy data and a poor choice of lifting functions. To prevent identifying an unstable model, the proposed approach imposes an asymptotic stability constraint on the Koopman model. The proposed method is formulated as a semidefinite program and its performance is compared to state-of-the-art methods with a simulated Duffing oscillator dataset and experimental soft robot dataset.
\end{abstract}

\section{Introduction} \noindent
Data-driven modeling is an attractive alternative to first-principles modeling, particularly when the underlying dynamics are complex or even unknown. Nonlinear system identification methods are needed when identifying a nonlinear model from data. The Koopman operator~\cite{Mezic2013, Williams2015, Mauroy2020, Korda2018} allows nonlinear dynamical systems to be represented as an infinite-dimensional linear system. However, in practice, the infinite-dimensional Koopman operator must be approximated as a finite-dimensional matrix when used for system identification, which is done by choosing a set of lifting functions that are used to identify the Koopman matrix. The lifting functions should be chosen such that the nonlinearities of the system are captured~\cite{Mezic2020, Budivsic2012}. Once lifting functions are chosen, and data is collected, the simplest means to identify the Koopman matrix is via linear least-squares. However, other methods, like Dynamic Mode Decomposition (DMD)~\cite{Proctor2016} and Extended DMD (EDMD)~\cite{Williams2015}, improve numerical conditioning when there are many lifting functions or many data snapshots, respectively.

When identifying a physical system, measurements are corrupted by sensor noise, which can introduce a bias in the identified model~\cite{Dawson2016, Hemati2017}. This bias is reflected by underestimated eigenvalues~\cite{Dawson2016, Hemati2017}, which result in larger decay rates in the biased system. The Koopman model identified using noisy data of a system with states and inputs has a bias in both its dynamics and input matrices. Although the literature provides methods to mitigate the impact of noise on the identification of the dynamics matrix of the approximated Koopman operator approximation~\cite{Dawson2016, Hemati2017, Haseli2019}, bias in the input matrix has not yet been addressed. Furthermore, a model identified from noisy measurements of an inherently asymptotically stable system can be unstable~\cite{Sinha2018}. A Koopman-based model can also become unstable due to a poor choice of lifting functions~\cite{Dahdah2022}.

Forward-backward DMD (fbDMD)~\cite{Dawson2016} provides a method that uses the forward- and backward-in-time dynamics of the system to obtain a dynamics matrix with reduced bias. This paper uses the underlying theory behind fbDMD to mitigate the impact of noise on the Koopman matrix obtained from EDMD with inputs, defined as fbEDMD. Additionally, to incorporate asymptotic stability into the problem, this work provides constraints, formulated as linear matrix inequalities (LMIs), that force the approximated Koopman system computed using fbEDMD to be asymptotically stable. The Koopman operator approximation problem, with bias reduction and asymptotic stability constraints, is then posed as a convex optimization problem with LMI constraints. The identification of the forward- and backward-in-time Koopman systems workflow and the proposed bias reduction method are summarized in Figure~\ref{summary_fig}.


\tikzstyle{overview} = [
    draw=gray,
    rounded corners=0.1cm,
    line width=1.5pt,
    minimum height=12cm,
    inner sep=0.1cm,
    anchor=north west,
    align=center,
]

\tikzstyle{block} = [
    draw,
    minimum width=2cm,
    minimum height=1.2cm
]

\tikzstyle{smallblock} = [
    draw,
    minimum width=1.2cm,
    minimum height=1.2cm
]

\tikzstyle{sum} = [
    draw,
    circle,
    minimum size=0.6cm,
]

\tikzstyle{diff} = [
    sum,
    label=170:{\small $+$},
    label=260:{\small $-$}
]

\tikzstyle{summ} = [
    sum,
    label=170:{\small $+$},
    label=80:{\small $+$}
]

\newsavebox{\arrow}
\begin{lrbox}{\arrow}
    \begin{tikzpicture}
        \draw[-latex, line width=2pt] (0cm, 0cm) -- (0cm, -1.0cm);
    \end{tikzpicture}
\end{lrbox}

\newsavebox{\rarrow}
\begin{lrbox}{\rarrow}
    \begin{tikzpicture}
        \node (origin) at (0,0) {};
        \draw[-latex, line width=2pt] (0cm, 1.0cm) -- (0.75cm, 1.0cm);
    \end{tikzpicture}
\end{lrbox}

\begin{figure}[t]
    \centering
    \resizebox{\textwidth}{!}{%
        \begin{tikzpicture}
            \node[
                overview,
                minimum width=8cm,
                label={north:(a) Collect, lift, and arrange snapshots},
            ] (step1) at (0, 0) {%
                \textit{Forward-in-time snapshots:}\\[1ex]
                \begin{overpic}[width=2.5cm]{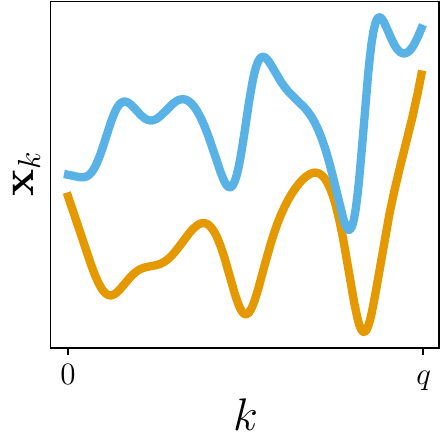}
                   \put(-10, -10){\includegraphics[width=2.5cm]{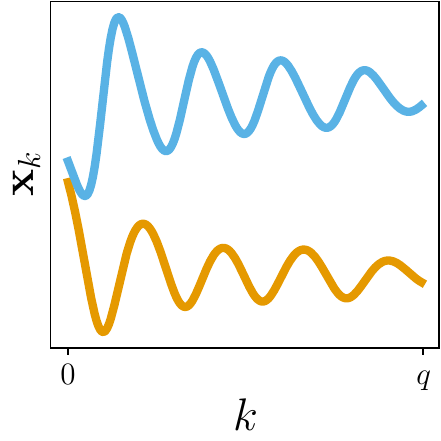}}
                   \put(-20, -20){\includegraphics[width=2.5cm]{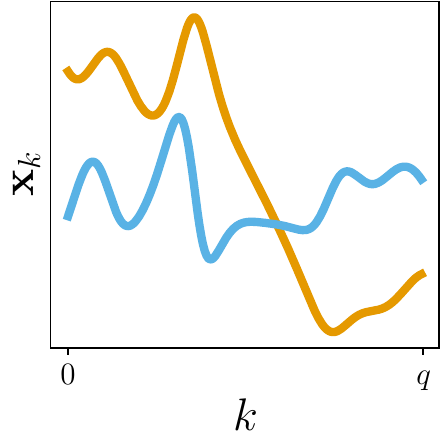}}
                \end{overpic}
                \usebox{\rarrow}\hspace{0.5cm}
                \begin{overpic}[width=2.5cm]{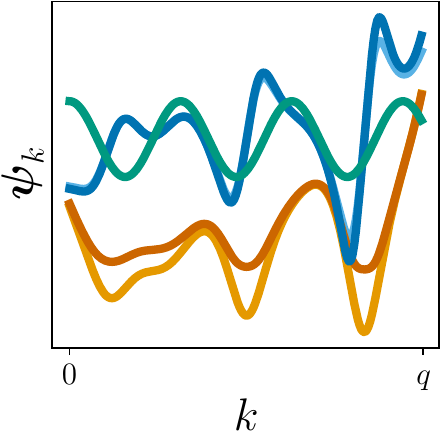}
                   \put(-10, -10){\includegraphics[width=2.5cm]{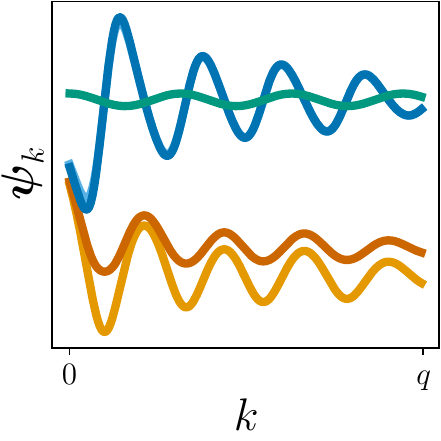}}
                   \put(-20, -20){\includegraphics[width=2.5cm]{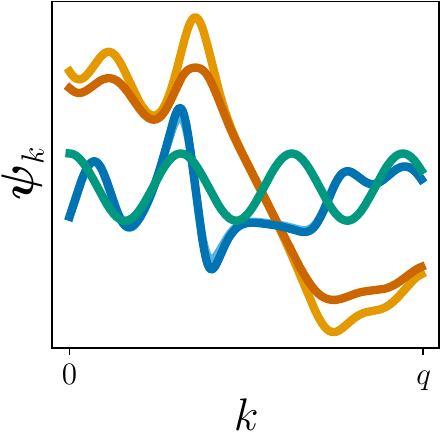}}
                \end{overpic}
                \\[3ex]
                $\mbs{\Psi} = \begin{bmatrix}
                    \mbs{\vartheta}_0 &
                    \mbs{\vartheta}_1 &
                    \cdots &
                    \mbs{\vartheta}_{q-1} \\
                    \mbs{\upsilon}_0 &
                    \mbs{\upsilon}_1 &
                    \cdots &
                    \mbs{\upsilon}_{q-1}
                \end{bmatrix},$\\[1ex]
                $\mbs{\Theta}_+ = \begin{bmatrix}
                    \mbs{\vartheta}_1 &
                    \mbs{\vartheta}_2 &
                    \cdots &
                    \mbs{\vartheta}_{q}
                \end{bmatrix}$
                \\[2ex]
                \textit{Backward-in-time snapshots:}\\[1ex]
                \begin{overpic}[width=2.5cm]{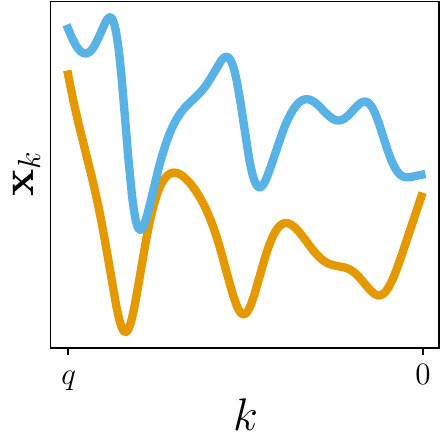}
                   \put(-10, -10){\includegraphics[width=2.5cm]{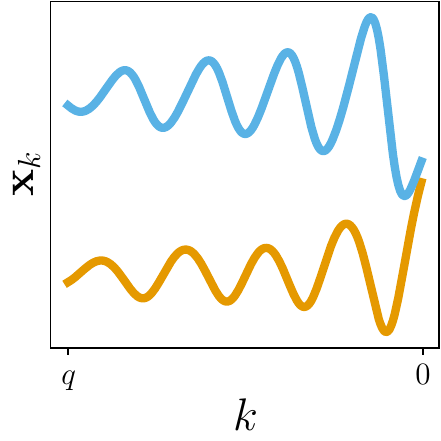}}
                   \put(-20, -20){\includegraphics[width=2.5cm]{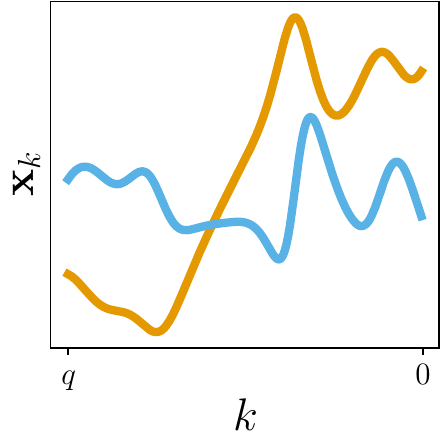}}
                \end{overpic}
                \usebox{\rarrow}\hspace{0.5cm}
                \begin{overpic}[width=2.5cm]{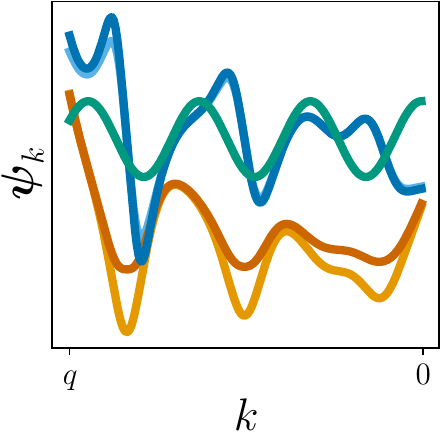}
                   \put(-10, -10){\includegraphics[width=2.5cm]{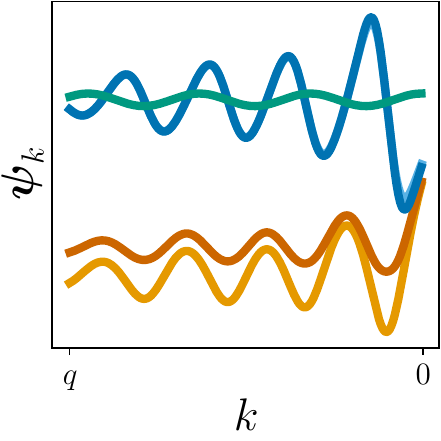}}
                   \put(-20, -20){\includegraphics[width=2.5cm]{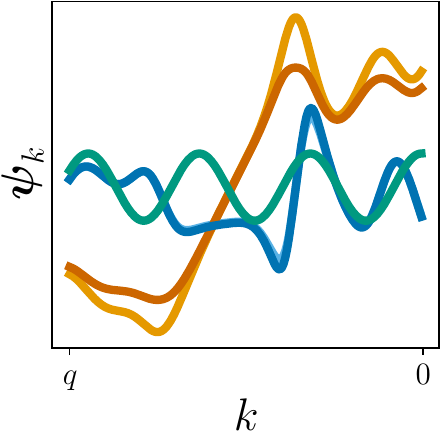}}
                \end{overpic}
                \\[3ex]
                $\mbs{\Theta} = \begin{bmatrix}
                    \mbs{\vartheta}_0 &
                    \mbs{\vartheta}_1 &
                    \cdots &
                    \mbs{\vartheta}_{q-1}
                \end{bmatrix}$\\[1ex]
                $\hat{\mbs{\Psi}} = \begin{bmatrix}
                    \mbs{\vartheta}_1 &
                    \mbs{\vartheta}_2 &
                    \cdots &
                    \mbs{\vartheta}_{q} \\
                    \mbs{\upsilon}_0 &
                    \mbs{\upsilon}_1 &
                    \cdots &
                    \mbs{\upsilon}_{q-1}
                \end{bmatrix}$
            };
            \node[
                overview,
                right=0.25cm of step1,
                minimum width=1cm,
                label={north:(b) Identify $\mbf{U}_\mathrm{ff}$ and $\mbf{U}_\mathrm{bb}$}
            ] (step2) {%
                \textit{Forward-in-time solution:}\\[1ex]
                $\underset{\mbf{U}_\mathrm{ff}}{\min}
                {\left\| \mbs{\Theta}_+ -
                \begin{bmatrix}
                    \mbf{A}_\mathrm{ff} &
                    \mbf{B}_\mathrm{ff}
                \end{bmatrix}
                \mbs{\Psi}
                \right\|}_\frob^2$
                \\
                $\mathrm{s.t.}\ \color{oivermillion}{\bar{\lambda}(\mbf{A}_\mathrm{ff}) < \bar{\rho}}$
                \\[3ex]
                \textit{Backward-in-time solution:}\\[1ex]
                $\underset{\mbf{U}_\mathrm{bb}}{\min}
                {\left\| \mbs{\Theta} -
                \begin{bmatrix}
                    \mbf{A}_\mathrm{bb} &
                    \mbf{B}_\mathrm{bb}
                \end{bmatrix}
                \hat{\mbs{\Psi}}
                \right\|}_\frob^2$
                \\
                $\mathrm{s.t.}\ \color{oiblue}{\underline{\lambda}(\mbf{A}_\mathrm{bb}) > \frac{1}{\bar{\rho}}}$
                \\[3ex]
                \includegraphics[width=5cm]{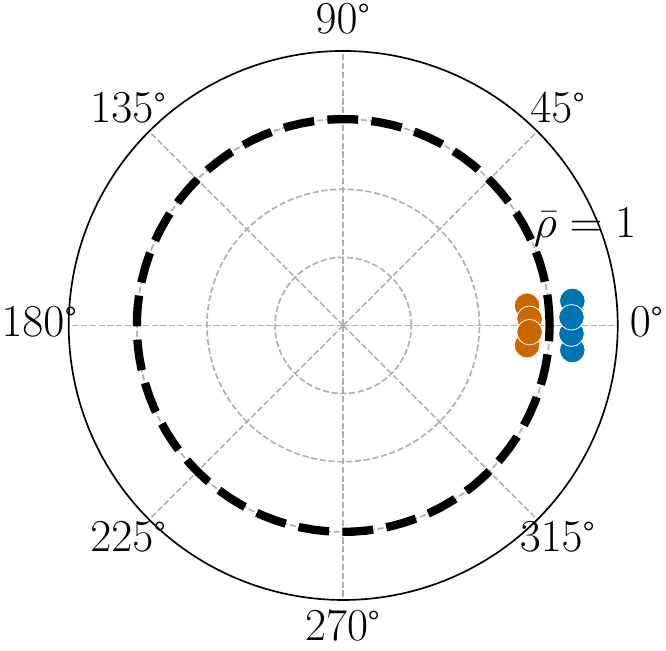}
            };
            \node[
                overview,
                right=0.25cm of step2,
                minimum width=1cm,
                label={north:(c) Compute unbiased $\tilde{\mbf{U}}$}
            ] (step3) {%
                \textit{Compute unbiased $\tilde{\mbf{A}}$:}\\[1ex]
                $\tilde{\mbf{A}} = \sqrt{\mbf{A}_\mathrm{ff} \mbf{A}_\mathrm{bb}^{-1}}$
                \\[3ex]
                \textit{Compute unbiased $\tilde{\mbf{B}}$:}\\[1ex]
                $\mbf{B}_\mathrm{fb} = -\mbf{A}_\mathrm{bb}^{-1} \mbf{B}_\mathrm{bb}$
                \\[1ex]
                $\tilde{\mbf{B}} = {(\mbf{1} + \tilde{\mbf{A}})}^\dagger(\mbf{B}_\mathrm{ff} + \mbf{A}_\mathrm{ff} \mbf{B}_\mathrm{fb})$
                \\[2ex]
                $\tilde{\mbf{U}} = \begin{bmatrix}
                    \tilde{\mbf{A}} & \tilde{\mbf{B}}
                \end{bmatrix}$
                \\[4ex]
                \includegraphics[width=4cm]{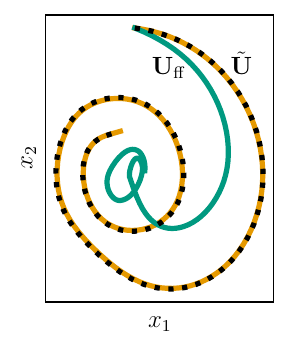}
            };
        \end{tikzpicture}
    }
    \caption{Identification process overview of the Koopman system with proposed bias reduction method and constraints, including the effects of the proposed method. $(a)$ First, the forward- and backward-in-time snapshot matrices are collected using noisy data from the system. $(b)$ Then, the forward- and backward-in-time Koopman matrices are approximated using asymptotic stability constraints. $(c)$ Finally, the Koopman matrix with reduced bias is computed. The proposed method identifies an asymptotically stable Koopman system representation with a reduced bias when noisy data is used.}  
    \label{summary_fig}
\end{figure}

\subsection{Related Work}
\subsubsection{Reducing the bias in the Koopman operator approximation}
\noindent
Methods in the literature provide different approaches to reduce the bias introduced by noise in a Koopman model. Forward-backward DMD~\cite{Dawson2016} leverages the forward- and backward-in-time dynamics of the system to produce a dynamics matrix with reduced bias. Another method is total least-squares DMD (TLS-DMD)~\cite{Golub1980, Hemati2017}. TLS-DMD addresses the introduction of bias in the DMD framework by assuming that noise is present in both snapshot matrices and uses a subspace projection step to obtain snapshot matrices with reduced noise. TLS-DMD generates snapshot matrices with reduced noise, but it is more computationally demanding than fbDMD~\cite{Dawson2016}. Although both fbDMD and TLS-DMD are methods that reduce the bias in a linear discrete-time system, they cannot be used for systems with inputs. In~\cite{Stevens2020}, the average of the forward- and backward-in-time input matrices, defined as their sum divided by two, is used to obtain a solution with reduced bias of the forward-in-time input matrix. Although this approach demonstrates good experimental results, it does not consider the expected value of the input matrix nor the impact of noise in the proposed solution for the input matrix with reduced bias. 

\subsubsection{Asymptotic stability of the Koopman operator approximation}
\noindent
When identifying an asymptotically stable system from noisy data, it is possible to identify an unstable model~\cite{Mamakoukas2020}. In~\cite{Dahdah2022}, the Koopman operator approximation problem is reformulated as a convex optimization problem with LMI constraints and uses bilinear matrix inequality (BMI) constraints to enforce asymptotic stability. Although~\cite{Dahdah2022} offers a modular way to enforce asymptotic stability on the forward-in-time Koopman operator approximation, it does not offer an equivalent solution for the backward-in-time dynamics, which is required in the proposed fbEDMD framework. In~\cite{Mabrok2023}, the BMI constraint used to impose asymptotic stability on the forward-in-time Koopman operator approximation is transformed into an LMI, allowing the optimization problem to be convex. This paper proposes a BMI constraint imposed on the backward-in-time dynamics, which is then transformed into an LMI formulation using \cite{Mabrok2023}, to obtain an asymptotically stable forward-in-time Koopman system with reduced bias. 

\subsection{Contributions}
\noindent
    The two main contributions of this paper are 1) the integration of inputs and 2) asymptotic stability constraints in the fbEDMD framework. The result is a forward-in-time Koopman system representation with reduced bias and is guaranteed to be asymptotically stable.

This paper shows that the proposed method, \emph{forward-backward EDMD with inputs and asymptotic stability constraints}, computes an asymptotically stable Koopman system with bias-reduced dynamics and input matrices when noisy data is used. The proposed method is then compared to Extended DMD~\cite{Williams2015}, Extended DMD with asymptotic stability constraint~\cite{Dahdah2022}, and Forward-Backward DMD with inputs using a simulated dataset of a Duffing oscillator and an experimental dataset of a soft robot arm.

\subsection{Organization}
\noindent
The remainder of this paper is as follows. Relevant theory is introduced in Section \ref{background}. The proposed expression of the Koopman matrix with reduced bias is presented in Section \ref{fbEDMD} and new BMI constraints to enforce asymptotic stability are presented in Section \ref{section_asym}. The optimization problem is posed in Section \ref{optimization_section} and results from the simulated and experimental datasets are shown in Section \ref{results_section} with concluding remarks in Section \ref{conclusion_section}.

\section{Background Theory} \label{background}

\subsection{Koopman operator theory}
\noindent
Consider the nonlinear, nonhomogenous, difference equation
\begin{equation} \label{disc_time_system}
    \mbf{x}_{k+1} = \mbf{f}(\mbf{x}_k, \mbf{u}_k),
\end{equation}
where $\mbf{x}_k \in \mathcal{M} \subseteq \mathbb{R}^{n \times 1}$, ${\mbf{u}_k \in \mathcal{N} \subseteq \mathbb{R}^{m \times 1}}$, and $\mbf{f}: \mathbb{R}^n \rightarrow \mathbb{R}^n$. Let a lifting function for this system be defined as ${\psi : \mathcal{M} \times \mathcal{N} \rightarrow \mathbb{R}}$, and the Koopman operator ${\mathcal{U} : \mathcal{H} \rightarrow \mathcal{H}}$, where $\mathcal{H}$ is an infinite-dimensional Hilbert space of lifting functions. Applying the Koopman operator to a lifting function advances that lifting function in time~\cite{Kutz2016, Dahdah2021, Dahdah2022}, such that  
\begin{equation}
    (\mathcal{U}\psi)(\mbf{x}_k, \mbf{u}_k) = \psi(\mbf{x}_{k+1}, \ast),
\end{equation}
where $\ast = \mbf{u}_k$ if the input has state-dependent dynamics. Otherwise, $\ast = 0$. 

Let the vector-valued lifting function $\mbs{\psi} : \mathcal{M} \times \mathcal{N} \rightarrow \mathbb{R}^{p \times 1}$ used to approximate the Koopman matrix be 
\begin{equation} \label{acte}
    \mbs{\psi}(\mbf{x}_k, \mbf{u}_k) = \begin{bmatrix}
        \mbs{\vartheta}(\mbf{x}_k) \\
        \mbs{\upsilon}(\mbf{x}_k, \mbf{u}_k)
    \end{bmatrix},
\end{equation}
where $\mbs{\vartheta} : \mathcal{M} \rightarrow \mathbb{R}^{p_\vartheta \times 1}$ and $\mbs{\upsilon} : \mathcal{M} \times \mathcal{N} \rightarrow \mathbb{R}^{p_\upsilon \times 1}$ such that $p_\vartheta + p_\upsilon = p$. Then, if the input has no state-dependent dynamics, the discrete-time system in \eqref{disc_time_system} can be expressed as
\begin{equation} \label{disc_time_koop}
    \mbs{\vartheta}(\mbf{x}_{k+1}) = \mbf{U}\mbs{\psi}(\mbf{x}_k, \mbf{u}_k) + \mbf{r}_k,
\end{equation}
where $\mbf{r}_k$ is the residual and the Koopman matrix is
\begin{equation} \label{koop}
    \mbf{U} = \begin{bmatrix}
        \mbf{A} & \mbf{B}
    \end{bmatrix}.
\end{equation}
The linear expression in \eqref{disc_time_koop} can then be rewritten as
\begin{equation}
    \mbs{\vartheta}(\mbf{x}_{k+1}) = \mbf{A}\mbs{\vartheta}(\mbf{x}_k) + \mbf{B}\mbs{\upsilon}(\mbf{x}_k, \mbf{u}_k) + \mbf{r}_k. \label{lin_eq_og}
\end{equation}
For the purpose of this paper, the lifting functions used to identify the input matrix \textbf{B} must be functions of $\mbf{u}_k$ only, such that \eqref{lin_eq_og} becomes
\begin{equation} \label{new_state_eq}
    \mbs{\vartheta}(\mbf{x}_{k+1}) = \mbf{A}\mbs{\vartheta}(\mbf{x}_k) + \mbf{B}\mbs{\upsilon}(\mbf{u}_k) + \mbf{r}_k, 
\end{equation}
and \eqref{acte} becomes
\begin{equation}
    \mbs{\psi}(\mbf{x}_k, \mbf{u}_k) = \begin{bmatrix}
        \mbs{\vartheta}(\mbf{x}_k) \\
        \mbs{\upsilon}(\mbf{u}_k)
    \end{bmatrix}.
\end{equation}
To express the approximation of the Koopman operator as part of a true linear system with input, output, and state, consider the output equation 
\begin{equation}\label{output_eq}
    \mbs{\zeta}_k = \mbf{C}\mbs{\vartheta}_k + \mbf{D}\mbs{\upsilon}_k + \mbf{v}_k,
\end{equation}
where $\mbf{v}_k$ is additive measurement noise and $\mbs{\zeta}_k \in \mathbb{R}^{p_\zeta \times 1}$ is the measured output at time step \emph{k}. In this paper, whenever a Koopman system is referenced, it is defined by \eqref{new_state_eq} and \eqref{output_eq} with $\mbf{C} = \textbf{1}$, where $\textbf{1}$ is the identity matrix, and $\mbf{D} = \textbf{0}$. As discussed in Section \ref{results_section}, artificial white noise is added to the datasets before lifting. Although it is assumed that the noise after lifting can be written as \eqref{output_eq}, the noise is originally introduced as 
\begin{equation}
    \mbf{y}_{k} = \mbf{C}\mbf{x}_k + \mbf{D}\mbf{u}_k + \mbs{\eta}_k,
\end{equation}
where $\mbf{y}_k$ and $\mbs{\eta}_k$ are the measured states and the added noise at timestep \emph{k}, respectively.

Let the snapshot matrices of the dataset $\mathcal{D} = \{\mbf{x}_k, \mbf{u}_k\}^q_{k=0}$ be expressed as
\begin{equation}
    \mbs{\Psi} = \begin{bmatrix}
        \mbs{\psi}_0 & \mbs{\psi}_1 & \cdots & \mbs{\psi}_{q-1}
    \end{bmatrix} \in \mathbb{R}^{p \times q}, \label{Psi} 
\end{equation}
and
\begin{equation}
    \mbs{\Theta}_+ = \begin{bmatrix}
        \mbs{\vartheta}_1 & \mbs{\vartheta}_2 & \cdots & \mbs{\vartheta}_q 
    \end{bmatrix} \in \mathbb{R}^{p_\vartheta \times q},
\end{equation}
where $\mbs{\psi}_k = \mbs{\psi}(\mbf{x}_k, \mbf{u}_k)$ and $\mbs{\vartheta}_k = \mbs{\vartheta}(\mbf{x}_k)$. The Koopman matrix can then be computed forward in time from the least-squares problem 
\begin{equation}
    \mbf{U}_\mathrm{f} = \argmin_{\mbf{U}} \;\;\; \left\|\mbs{\Theta}_+ - \mbf{U}\mbs{\Psi}\right\|^2_\frob, \label{vanilla_least}
\end{equation}
whose solution is~\cite{Kutz2016} 
\begin{equation} \label{true_edmd}
    \mbf{U}_{\mathrm{f}} = \mbs{\Theta}_+\mbs{\Psi}^\dagger, 
\end{equation}
where $\|\cdot\|_\frob$ represents the Frobenius norm, $(\cdot)^\dagger$ is the Moore-Penrose pseudoinverse, and $\mbs{\Psi}$ is assumed to be full row rank.

To reduce the numerical challenges when computing the pseudoinverse of $\mbs{\Psi}$ with many snapshots, \eqref{true_edmd} can be  rewritten using EDMD~\cite{Williams2015}, such that
\begin{equation} \label{forw_edmd1}
    \mbf{U}_\mathrm{f} = \mbs{\Theta}_+\mbs{\Psi}^\dagger = \left(\mbs{\Theta}_+\mbs{\Psi}^\trans\right)\left(\mbs{\Psi}\mbs{\Psi}^\trans\right)^\dagger = \mbf{G}_\mathrm{f}\mbf{H}_\mathrm{f}^\dagger, 
\end{equation}
where 
\begin{equation} \label{forw_edmd}
    \mbf{G}_\mathrm{f} = \frac{1}{q}\mbs{\Theta}_+\mbs{\Psi}^\trans \;\; \in \mathbb{R}^{p_\vartheta \times p}, \;\;\;\; \mbf{H}_\mathrm{f} = \frac{1}{q}\mbs{\Psi}\mbs{\Psi}^\trans \;\; \in \mathbb{R}^{p \times p},
\end{equation}
and \emph{q} is the number of snapshots. 

\subsection{Backward-in-time EDMD}
\noindent
A similar expression as \eqref{forw_edmd1} for the Koopman matrix can also be defined using the backward-in-time dynamics of a system. This is achieved by modifying the snapshot matrices of the dataset $\mathcal{D} = \{\mbf{x}_k, \mbf{u}_k\}^q_{k=0}$ as
\begin{align}
    \hat{\mbs{\Psi}} &= \begin{bmatrix}
        \mbs{\Theta}_+ \\
        \mbs{\Upsilon}
    \end{bmatrix} = \begin{bmatrix}
        \mbs{\vartheta}_1 & \mbs{\vartheta}_2 & \cdots & \mbs{\vartheta}_{q} \\
        \mbs{\upsilon}_0 & \mbs{\upsilon}_1 & \cdots & \mbs{\upsilon}_{q-1}
    \end{bmatrix} \in \mathbb{R}^{p \times q}, \label{back_edmd_1} \\
    \mbs{\Theta} &= \begin{bmatrix}
        \mbs{\vartheta}_0 & \mbs{\vartheta}_1 & \cdots & \mbs{\vartheta}_{q-1}
    \end{bmatrix}\;\; \in \mathbb{R}^{p_\vartheta \times q}. \label{back_edmd_2}
\end{align}
The backward Koopman matrix $\mbf{U}_\mathrm{b}$ propagates the snapshot matrix $\hat{\mbs{\Psi}}$ backward by one time step, such that 
\begin{equation}
    \mbf{U}_\mathrm{b}\hat{\mbs{\Psi}} = \mbs{\Theta}.
\end{equation}
The EDMD solution to the backward-in-time Koopman matrix is then 
\begin{equation} \label{back_edmd1}
    \mbf{U}_\mathrm{b} = \mbs{\Theta}\hat{\mbs{\Psi}}^\dagger = \left(\mbs{\Theta}\hat{\mbs{\Psi}}^\trans\right)\left(\hat{\mbs{\Psi}}\hat{\mbs{\Psi}}^\trans\right)^\dagger =
    \mbf{G}_\mathrm{b}{\mbf{H}_\mathrm{b}}^\dagger, 
\end{equation}
where
\begin{equation}
    \mbf{G}_\mathrm{b} = \frac{1}{q}\mbs{\Theta}\hat{\mbs{\Psi}}^\trans \;\; \in \mathbb{R}^{p_\vartheta \times p} \;\;\;\;\; \text{and} \;\;\;\;\; \mbf{H}_\mathrm{b} = \frac{1}{q}\hat{\mbs{\Psi}}\hat{\mbs{\Psi}}^\trans \;\; \in \mathbb{R}^{p \times p}. \label{back_edmd}
\end{equation}

\section{Forward-backward EDMD} \label{fbEDMD}
\noindent
Forward-backward EDMD is a method that uses both the forward- and backward-in-time dynamics to approximately cancel a bias introduced by the presence of noise when identifying the dynamics matrix~\cite{Dawson2016}. This paper extends fbDMD, which computes a dynamics matrix with reduced bias using DMD, to fbEDMD with inputs, which uses EDMD to compute both the dynamics and input matrices with reduced bias. In this section, mathematical expressions are developed to relate the forward- and backward-in-time dynamics. Then, an expression to calculate the input matrix with reduced bias is derived, following the same reasoning as~\cite{Dawson2016}.

\subsection{Forward and backward state-space matrices}
\noindent
To relate the forward- and backward-in-time dynamics to each other, both state-space models are expressed as a function of each other. In this paper, similarly to \cite{Dawson2016}, to allow the forward-in-time dynamics to be expressed using the backward-in-time dynamics and vice versa, the dynamics of the system are assumed to be invertible, such that $\mbf{A}_\mathrm{ff}^{-1}$ exists. Let the forward-in-time dynamics be 
\begin{equation} \label{forw_ss}
    \mbf{x}_k = \mbf{A}_\mathrm{ff}\mbf{x}_{k-1} + \mbf{B}_\mathrm{ff}\mbf{u}_{k-1}, 
\end{equation}
where $\mbf{A}_\mathrm{ff}$ and $\mbf{B}_\mathrm{ff}$ are the dynamics and input matrices with subscript notation as follows. The first letter in the subscript indicates the direction in time of the dynamics described by the matrix. The second letter indicates whether the forward or backward snapshot matrices were used to compute the matrix. Thus, in the case of \eqref{forw_ss}, the subscripts indicate the forward dynamics are considered, and the forward snapshot matrices are used to compute the matrices.

Let the backward-in-time dynamics, identified using the backward-in-time dataset, be~\cite{juang1994}
\begin{equation}
     \mbf{x}_{k-1} = \mbf{A}_\mathrm{bb}\mbf{x}_{k} + \mbf{B}_\mathrm{bb}\mbf{u}_{k-1}, \label{back_ss}
\end{equation}
where the subscripts indicate the backward dynamics are considered, and backward snapshot matrices are used to compute the matrices. Assuming that the dynamics of the system are invertible, the backward-in-time dynamics can be defined using the forward-in-time dynamics described by \eqref{forw_ss} as
\begin{align}
    \mbf{x}_{k-1} &= \left(\mbf{A}_\mathrm{ff}\right)^{-1}(\mbf{x}_k - \mbf{B}_\mathrm{ff}\mbf{u}_{k-1}) \\
    &= \left(\mbf{A}_\mathrm{ff}\right)^{-1}\mbf{x}_k - \left(\mbf{A}_\mathrm{ff}\right)^{-1}\mbf{B}_\mathrm{ff}\mbf{u}_{k-1} \\
    &= \mbf{A}_\mathrm{bf}\mbf{x}_{k} + \mbf{B}_\mathrm{bf}\mbf{u}_{k-1},
\end{align}
leading to the relationships
\begin{align}
    &\mbf{A}_\mathrm{fb} = \mbf{A}_\mathrm{bb}^{-1}, \label{iden_1}\\ 
    &\mbf{B}_\mathrm{fb} = -\mbf{A}_\mathrm{bb}^{-1}\mbf{B}_\mathrm{bb}, \label{iden_2}\\
    &\mbf{A}_\mathrm{bf} = \mbf{A}_\mathrm{ff}^{-1}, \label{iden_3}\\
    &\mbf{B}_\mathrm{bf} = -\mbf{A}_\mathrm{ff}^{-1}\mbf{B}_\mathrm{ff}. \label{iden_4}
\end{align}
Note that when noise is present in the dataset used to identify the dynamics and input matrices, \eqref{iden_1}--\eqref{iden_4} hold approximately. From the relationships developed in \eqref{iden_1}--\eqref{iden_4}, the true dynamics matrix  $\bar{\mbf{A}}$ and true input matrix $\bar{\mbf{B}}$, which are expressed forward in time, can be used to obtain the true backward-in-time matrices as 
\begin{align}
    &\bar{\mbf{A}}_\mathrm{b} = \bar{\mbf{A}}^{-1}, \label{true_iden_3}\\
    &\bar{\mbf{B}}_\mathrm{b} = -\bar{\mbf{A}}^{-1}\bar{\mbf{B}}. \label{true_iden_4}
\end{align}

\subsection{Reducing the bias in the Koopman matrix}
\noindent
Let the measured lifted snapshot matrix $\mbs{\Psi}$ be written as 
\begin{equation} \label{param_noise}
     \mbs{\Psi} = \bar{\mbs{\Psi}} + \mbs{N}_{\mbs{\Psi}}
     = \begin{bmatrix}
        \bar{\mbs{\Theta}} \\
        \bar{\mbs{\Upsilon}}
    \end{bmatrix} + \begin{bmatrix}
            \mbs{N}_{\mbs{\vartheta}} \\
            \mbs{0}
        \end{bmatrix},
\end{equation}
where $\bar{\mbs{\Psi}}$ is the true snapshot matrix, $\mbs{N}_{\mbs{\Psi}}$ is the noise within the measured lifted snapshot matrix, which is assumed to be additive~\cite{Dawson2016, Pintelon2012}, and $\bar{\mbs{\Upsilon}}$ is the true input-dependent block of the snapshot matrix. The noise is assumed to be independent and identically distributed with zero mean and finite variance, as sensor noise is often modeled in this manner~\cite[\S 6.7.3.1]{Pintelon2012}. This assumption is made when the noise is initially added to the dataset and after the lifting step. Although this assumption does not always hold, as demonstrated in Section \ref{results_section}, results using this assumption are satisfactory in the examples considered. Additionally, the inputs in \eqref{param_noise} are assumed to be known exactly~\cite[\S 7.7.3.2]{Pintelon2012} as opposed to the state-dependent block of the snapshot matrix.

As derived in~\cite{Dawson2016}, the expected value of the biased Koopman matrix is
\begin{equation} \label{koopman_expand}
    E\left[\mbf{U}\right] \approx \bar{\mbf{U}}\left(\textbf{1} - E\left[\mbs{N}_{\mbs{\Psi}}\mbs{N}_{\mbs{\Psi}}^\trans\right]\left(\mbs{\Psi}\mbs{\Psi}^\trans\right)^{-1}\right),
\end{equation}
where $\bar{\mbf{U}}$ is the true Koopman matrix.
Substituting \eqref{koop} into \eqref{koopman_expand} results in 
\begin{align}
    E\left[\mbf{U}\right] &\approx \begin{bmatrix}
        \bar{\mbf{A}} & \bar{\mbf{B}}
    \end{bmatrix}\left(\textbf{1} - E\left[\mbs{N}_{\mbs{\Psi}}\mbs{N}_{\mbs{\Psi}}^\trans\right]\left(\mbs{\Psi}\mbs{\Psi}^\trans\right)^{-1}\right) \\
    &= \begin{bmatrix}
         \bar{\mbf{A}} & \bar{\mbf{B}}
    \end{bmatrix} -
    \begin{bmatrix}
        \bar{\mbf{A}}\mbf{R} & \mbs{0}
        \end{bmatrix}
    \begin{bmatrix}
        \mbs{\Theta}\mbs{\Theta}^\trans & \mbs{\Theta}\mbs{\Upsilon}^\trans \\
        \star & \mbs{\Upsilon}\mbs{\Upsilon}^\trans
    \end{bmatrix}
    ^{-1}, \label{koop_deriv_1}
\end{align}
where $\mbf{R} = E\left[\mbs{N}_{\mbs{\vartheta}}\mbs{N}_{\mbs{\vartheta}}^\trans\right]$ and $\star$ defines the symmetric part of the matrix. Then, using the block inversion theorem~\cite{Lu2002}, \eqref{koop_deriv_1} becomes 
\begin{align}
     E\left[\mbf{U}\right] &\approx \begin{bmatrix}
         \bar{\mbf{A}} & \bar{\mbf{B}}
    \end{bmatrix} -
    \begin{bmatrix}
    \bar{\mbf{A}}\mbf{R} & \mbs{0}
        \end{bmatrix}
    \begin{bmatrix}
        \mbf{J}_{11} & \mbf{J}_{12} \\
        \star & \mbf{J}_{22}
    \end{bmatrix} \\
    &= \begin{bmatrix} \bar{\mbf{A}}\left(\textbf{1} - \mbf{R}\mbf{J}_{11}\right) & \bar{\mbf{B}} - \bar{\mbf{A}}\mbf{R}\mbf{J}_{12}
        \end{bmatrix},
\end{align}
which implies that
\begin{align}
    E\left[\mbf{A}\right] &= \bar{\mbf{A}}\left(\textbf{1} - \mbf{R}\mbf{J}_{11}\right), \label{A_expected}\\
    E\left[\mbf{B}\right] &= \bar{\mbf{B}} - \bar{\mbf{A}}\mbf{R}\mbf{J}_{12}. \label{B_expected}
\end{align}
Note that the recovered expected value of the dynamics matrix in \eqref{A_expected} is the same solution as in~\cite{Dawson2016}. Additionally, both \eqref{A_expected} and \eqref{B_expected} hold when applied forward and backward in time, where the forward-in-time expressions are 
\begin{align}
     E\left[\mbf{A}_\mathrm{ff}\right] &= \bar{\mbf{A}}\left(\textbf{1} - \mbf{R}\mbf{J}_{11}\right), \label{A_expected_forw}\\
    E\left[\mbf{B}_\mathrm{ff}\right] &= \bar{\mbf{B}} - \bar{\mbf{A}}\mbf{R}\mbf{J}_{12}, \label{B_expected_forw}   
\end{align}
and the backward-in-time expressions are
\begin{align}
     E\left[\mbf{A}_\mathrm{bb}\right] &= \bar{\mbf{A}}^{-1}\left(\textbf{1} - \mbf{R}\mbf{J}_{11}\right), \label{A_expected_back}\\
    E\left[\mbf{B}_\mathrm{bb}\right] &= -\bar{\mbf{A}}^{-1}\bar{\mbf{B}} - \bar{\mbf{A}}^{-1}\mbf{R}\mbf{J}_{12}, \label{B_expected_back}   
\end{align}
where the identities in \eqref{true_iden_3} and \eqref{true_iden_4} are used in \eqref{A_expected_back} and \eqref{B_expected_back}, respectively. Following this result,~\cite{Dawson2016} shows that one way to obtain the dynamics matrix with reduced bias is
\begin{equation}
    \tilde{\mbf{A}} \approx \sqrt{\mbf{A}_\mathrm{ff}\mbf{A}_\mathrm{fb}}, \label{paper_unbiased_A}
\end{equation}
where $\sqrt{\cdot}$ is the matrix square root computed using a Schur decomposition \cite{Deadman2012} and $\tilde{\mbf{A}}$ is the forward-in-time dynamics matrix with reduced bias. To obtain \eqref{paper_unbiased_A}, consider the approximate expression 
\begin{align}
    \tilde{\mbf{A}} &\approx \sqrt{E\left[\mbf{A}_\mathrm{ff}\mbf{A}_\mathrm{bb}^{-1}\right]} \label{Dawson_1} \\
    & \approx \sqrt{E\left[\mbf{A}_\mathrm{ff}\right]E\left[\mbf{A}_\mathrm{bb}^{-1}\right]} \label{Dawson_2} \\
    &\approx \sqrt{E\left[\mbf{A}_\mathrm{ff}\right]E\left[\mbf{A}_\mathrm{bb}\right]^{-1}}, \label{Dawson_3}
\end{align}
    where \eqref{Dawson_2} and \eqref{Dawson_3} are obtained by assuming that the distributive property with the expected value operator holds for multiplications and exponents, respectively. Note that \eqref{paper_unbiased_A} can be recovered from \eqref{Dawson_1} by using \eqref{iden_1}. Then, substituting \eqref{A_expected_forw} and \eqref{A_expected_back} into \eqref{Dawson_3} leads to 
\begin{align}
   \tilde{\mbf{A}} & \approx \sqrt{\bar{\mbf{A}}\left(\textbf{1} - \mbf{R}\mbf{J}_{11}\right)\left(\textbf{1} - \mbf{R}\mbf{J}_{11}\right)^{-1}\bar{\mbf{A}}} \label{Dawson_4} \\ 
    &= \bar{\mbf{A}}, \label{Dawson_5}
\end{align}
where $E\left[\mbf{A}_\mathrm{bb}\right]$ is assumed to be invertible.

The solution for the input matrix with reduced bias $\tilde{\mbf{B}}$ is found by noticing that the noise term in \eqref{B_expected} is additive to $\bar{\mbf{B}}$. Consider the expression
\begin{align}
    E\left[\mbf{B}_{\mathrm{ff}} + \mbf{A}_\mathrm{ff}\mbf{B}_{\mathrm{fb}}\right] & = E\left[\mbf{B}_{\mathrm{ff}}\right] + E\left[\mbf{A}_\mathrm{ff}\mbf{B}_{\mathrm{fb}}\right] \\
    &= E\left[\mbf{B}_{\mathrm{ff}}\right] + E\left[-\mbf{A}_\mathrm{ff}\mbf{A}_{\mathrm{bb}}^{-1}\mbf{B}_{\mathrm{bb}}\right], \label{lortie_proof_1}
\end{align}
where ${\mbf{B}_\mathrm{fb} = -\mbf{A}_\mathrm{bb}^{-1}\mbf{B}_\mathrm{bb}}$ from \eqref{iden_2}. It will now be shown that \eqref{lortie_proof_1} results in an estimate of the input matrix with reduced bias. Under the assumption that the distributive property for exponents and multiplications hold for the expected value operator~\cite{Dawson2016}, \eqref{lortie_proof_1} can be rewritten as 
\begin{equation}
   E\left[\mbf{B}_{\mathrm{ff}} + \mbf{A}_\mathrm{ff}\mbf{B}_{\mathrm{fb}}\right] \approx E\left[\mbf{B}_{\mathrm{ff}}\right] - E\left[\mbf{A}_\mathrm{ff}\right]E\left[\mbf{A}_{\mathrm{bb}}\right]^{-1}E\left[\mbf{B}_{\mathrm{bb}}\right]. \label{lortie_proof_2}
\end{equation}
Next, substituting \eqref{A_expected_forw}--\eqref{B_expected_back} in \eqref{lortie_proof_2} gives  
\begin{align}
     E\left[\mbf{B}_{\mathrm{ff}} +  \mbf{A}_\mathrm{ff}\mbf{B}_{\mathrm{fb}}\right] &\approx \bar{\mbf{B}} - \bar{\mbf{A}}\mbf{R}\mbf{J}_{12} - \bar{\mbf{A}}\left(\textbf{1} - \mbf{R}\mbf{J}_{11}\right)\left(\bar{\mbf{A}}^{-1}\left(\textbf{1} - \mbf{R}\mbf{J}_{11}\right)\right)^{-1} \left(-\bar{\mbf{A}}^{-1}\bar{\mbf{B}} - \bar{\mbf{A}}^{-1}\mbf{R}\mbf{J}_{12}\right) \label{lortie_proof_3.1} \\
      & = \bar{\mbf{B}} - \bar{\mbf{A}}\mbf{R}\mbf{J}_{12} + \bar{\mbf{A}}\bar{\mbf{B}} + \bar{\mbf{A}}\mbf{R}\mbf{J}_{12} \label{lortie_proof_3.2} \\
    & = \bar{\mbf{B}} + \bar{\mbf{A}}\bar{\mbf{B}}. \label{lortie_proof_4}
\end{align}
Notice that the bias terms in \eqref{lortie_proof_3.1} and \eqref{lortie_proof_3.2} cancel out to create \eqref{lortie_proof_4}. Since the true dynamics and input matrices in \eqref{lortie_proof_4} can't be exactly computed, let them be approximated by the dynamics and input matrices with reduced bias, such that $\bar{\mbf{A}} \approx \tilde{\mbf{A}}$ and $\bar{\mbf{B}} \approx \tilde{\mbf{B}}$. Additionally, let the expected value term be approximately equal to the operand of the expected value operator, then the solution for the forward-in-time input matrix with reduced bias is found by rewriting \eqref{lortie_proof_4}, such that
\begin{align}
   \tilde{\mbf{B}} + \tilde{\mbf{A}}\tilde{\mbf{B}} &\approx \mbf{B}_{\mathrm{ff}} + \mbf{A}_\mathrm{ff}\mbf{B}_{\mathrm{fb}}, \\
    \tilde{\mbf{B}} &\approx (\textbf{1} + \tilde{\mbf{A}})^{\dagger}(\mbf{B}_{\mathrm{ff}} + \mbf{A}_\mathrm{ff}\mbf{B}_{\mathrm{fb}}). \label{avg_B}
\end{align}
In summary, fbEDMD with inputs is presented in this section. It is first shown in \eqref{paper_unbiased_A} that the forward-in-time dynamics matrix with reduced bias $\tilde{\mbf{A}}$ is computed using the forward- and backward-in-time dynamics matrices, as is done in~\cite{Dawson2016}. Then, inspired by \cite{Dawson2016}, a solution for the forward-in-time input matrix with reduced bias $\tilde{\mbf{B}}$ is derived in \eqref{avg_B}. Finally, the Koopman matrix with reduced bias is constructed as
\begin{equation}
\tilde{\mbf{U}} = \begin{bmatrix}
    \tilde{\mbf{A}} & \tilde{\mbf{B}}
\end{bmatrix}.
\end{equation}

\section{Asymptotic stability of the Koopman system} \label{section_asym}
\noindent
The forward-in-time Koopman matrix $\mbf{U}_{\mathrm{ff}}$ is found by solving the least-squares problem~\cite{Dahdah2022}
\begin{equation}
  \mbf{U}_{\mathrm{ff}} = \argmin_{\mbf{A}_{\mathrm{ff}}, \mbf{B}_{\mathrm{ff}}} \;\;\;\; \left\|\mbs{\Theta}_+ - \begin{bmatrix}
        \mbf{A}_{\mathrm{ff}} & \mbf{B}_{\mathrm{ff}}
    \end{bmatrix}\mbs{\Psi}\right\|^2_\frob. \label{vanilla_cost_1}
\end{equation}
The optimization problem minimizes the cost function in \eqref{vanilla_cost_1} such that $\mbf{A}_{\mathrm{ff}}$ and $\mbf{B}_{\mathrm{ff}}$ are computed. Similarly, the backward-in-time Koopman matrix approximation $\mbf{U}_\mathrm{bb}$ is computed by solving
\begin{equation}
  \mbf{U}_\mathrm{bb} = \argmin_{\mbf{A}_{\mathrm{bb}}, \mbf{B}_{\mathrm{bb}}} \;\;\;\; \left\|\mbs{\Theta} - \begin{bmatrix}
        \mbf{A}_{\mathrm{bb}} & \mbf{B}_{\mathrm{bb}}
    \end{bmatrix}\hat{\mbs{\Psi}}\right\|^2_\frob, \label{vanilla_cost_2}
\end{equation}
where $\hat{\mbs{\Psi}}$ and $\mbs{\Theta}$ are defined by \eqref{back_edmd_1} and \eqref{back_edmd_2}, respectively.

When identifying an asymptotically stable system, it is important that the data-driven model is also asymptotically stable~\cite{Mamakoukas2020}. The identified Koopman system can be unstable if the lifting functions are poorly chosen~\cite{Dahdah2022} or if noise is present in the data used to construct the snapshot matrices. 

This section establishes a method to enforce asymptotic stability in the Koopman system with reduced bias by constraining the spectral radius of its dynamics matrix to be bounded above by the constant $\bar{\rho}$. 

\subsection{Constraint for forward-in-time dynamics}
\noindent
In forward time, the eigenvalues of $\mbf{A}_\mathrm{ff}$ must be strictly within the unit circle for the system to be asymptotically stable. To ensure this is achieved, the Lyapunov constraint~\cite[\S 2.14]{Caverly2019}
\begin{equation} \label{forw_lyap}
\mbf{A}_\mathrm{ff}\mbf{P}\mbf{A}_\mathrm{ff}^\trans - \bar{\rho}^2\mbf{P} < 0,
\end{equation}
with $\mbf{P} > 0$, is used to force the largest eigenvalue of $\mbf{A}_\mathrm{ff}$ to be bounded from above by $\bar{\rho}$, where $\bar{\rho}$ is usually $0 < \bar{\rho} < 1$. Note that throughout this paper, negative definite and negative semidefinite matrices are denoted as $< 0$ and $\le 0$, respectively. Using the Schur complement, \eqref{forw_lyap} can be rewritten as~\cite{Dahdah2022, Caverly2019}
\begin{equation} \label{cons_asym_forw}
    \begin{bmatrix}
        \Bar{\rho}\mbf{P} & \mbf{A}_\mathrm{ff}\mbf{P} \\
        \star & \Bar{\rho}\mbf{P}
    \end{bmatrix} > 0.
\end{equation}
Equation \eqref{cons_asym_forw} is a BMI due to the product of unknowns $\mbf{A}_\mathrm{ff}\mbf{P}$.

\subsection{Constraint for backward-in-time dynamics}
\noindent
Similarly, to ensure that a similar Lyapunov constraint to \eqref{forw_lyap} enforces asymptotic stability for the forward-in-time system identified with the backward-in-time snapshot matrices, a congruence transformation is done on \eqref{forw_lyap}. Premultiplying and postmultiplying by ${\mbf{A}_\mathrm{ff}^{-1}}$ and ${\mbf{A}_\mathrm{ff}^{-\trans}}$ \eqref{forw_lyap}, respectively, gives
\begin{equation} \mbf{P}- \bar{\rho}^2\mbf{A}_\mathrm{ff}^{-1}\mbf{P}{\mbf{A}_\mathrm{ff}^{-\trans}} < 0. \label{here82}
\end{equation}
Negating \eqref{here82} and replacing $\mbf{A}_\mathrm{ff}^{-1}$ with $\mbf{A}_\mathrm{bf}$ using \eqref{iden_3} gives
\begin{equation}
\bar{\rho}^2\mbf{A}_\mathrm{bf}\mbf{P}\mbf{A}_\mathrm{bf}^\trans - \mbf{P} > 0.\label{deriv_1}
\end{equation}
In practice, \eqref{deriv_1} only uses data collected backwards in time, meaning substituting $\mbf{A}_{\mathrm{bf}}$ with $\mbf{A}_{\mathrm{bb}}$ in \eqref{deriv_1}, results in 
\begin{equation}
    \mbf{A}_\mathrm{bb}\mbf{P}\mbf{A}_\mathrm{bb}^\trans - \frac{1}{\bar{\rho}^2}\mbf{P} > 0, \label{here_43}
\end{equation} 
where $\mbf{P} > 0$. It is shown in the \nameref{Appendix} that the Lyapunov constraint in \eqref{here_43} is equivalent to bounding the eigenvalues of $\mbf{A}_\mathrm{bb}$ from below by $\frac{1}{\bar{\rho}}$. Let \eqref{here_43} be rewritten as
\begin{equation}
    (\bar{\rho}\mbf{P}^\trans\mbf{A}_\mathrm{bb}^\trans)^\trans\mbf{P}^{-1}(\bar{\rho}\mbf{P}^\trans\mbf{A}_\mathrm{bb}^\trans) - \mbf{P} > 0.   \label{young start}
\end{equation}
Next, consider a special case of Young's inequality \cite{Caverly2019},
\begin{equation}
    \mbf{G}^\trans\mbf{S}^{-1}\mbf{G} \ge \mbf{G} + \mbf{G}^\trans - \mbf{S}, \label{young special}
\end{equation}
where $\mbf{G} \in \mathbb{R}^{n\times n}$, $\mbf{S} > 0$. Letting $\mbf{G} = \bar{\rho}\mbf{P}^\trans\mbf{A}_\mathrm{bb}^\trans$ and $\mbf{S} = \mbf{P}$, it follows that \eqref{young special} can be used with \eqref{young start} as
\begin{align}
    \mbf{G}^\trans\mbf{S}^{-1}\mbf{G} - \mbf{S} &\ge \mbf{G} + \mbf{G}^\trans - \mbf{S} - \mbf{S}  \\
    &= \bar{\rho}\mbf{P}^\trans\mbf{A}_\mathrm{bb}^\trans + \bar{\rho}\mbf{A}_\mathrm{bb}\mbf{P} - 2\mbf{P}. \label{mod lyap}
\end{align}
Then, the modified Lyapunov constraint described by \eqref{mod lyap} can be constrained as 
\begin{equation}
    \bar{\rho}\mbf{P}^\trans\mbf{A}_\mathrm{bb}^\trans + \bar{\rho}\mbf{A}_\mathrm{bb}\mbf{P} - 2\mbf{P} > 0, \label{back lyap}
\end{equation}
which ensures \eqref{here_43} holds sufficiently. The BMI constraint presented in \eqref{back lyap} ensures that the forward-in-time model identified using the backward-in-time snapshot matrices is asymptotically stable.

\subsection{Asymptotic stability of the system with reduced bias} \label{stability_proof}
\noindent
In this section, it is shown that it is sufficient for \eqref{forw_lyap} and \eqref{here_43} to hold for the spectral radius of $\sqrt{\mbf{A}_\mathrm{ff}\mbf{A}_\mathrm{fb}}$ to be bounded above by $\bar{\rho}$. First, $\mbf{A}_\mathrm{ff}\mbf{A}_\mathrm{fb}$ is shown to have a spectral radius bounded above by $\bar{\rho}^2$, then the matrix square root operator is shown to conserve the desired bounds given to the eigenvalues of  $\mbf{A}_\mathrm{ff}\mbf{A}_\mathrm{fb}$.

It will now be shown that, when \eqref{forw_lyap} and \eqref{here_43} hold with $\mbf{P} > 0$,
\begin{equation} \label{lyap_avg}
    (\mbf{A}_\mathrm{ff}\mbf{A}_\mathrm{fb})^\trans \mbf{P} \mbf{A}_\mathrm{ff}\mbf{A}_\mathrm{fb} - \bar{\rho}^4\mbf{P} < 0.
\end{equation}
The constraints in \eqref{lyap_avg} forces the eigenvalues of $\mbf{A}_\mathrm{ff}\mbf{A}_\mathrm{fb}$ to be bounded above by $\bar{\rho}^2$ \cite{Caverly2019}. To show this statement by contraposition, consider the contrapositive of \eqref{lyap_avg},
\begin{equation}
    \mbf{A}_\mathrm{fb}^\trans\mbf{A}_\mathrm{ff}^\trans \mbf{P} \mbf{A}_\mathrm{ff}\mbf{A}_\mathrm{fb} - \bar{\rho}^4\mbf{P} \ge 0. \label{deriv_2}
\end{equation}
Then, applying a congruence transform on \eqref{deriv_2} by premultiplying by $\mbf{A}_\mathrm{fb}^{-\trans}$ and postmultiplying by $\mbf{A}_\mathrm{fb}^{-1}$ gives
\begin{align}
   \mbf{A}_\mathrm{ff}^\trans \mbf{P} \mbf{A}_\mathrm{ff} - \bar{\rho}^4\mbf{A}_\mathrm{fb}^{-\trans}\mbf{P}\mbf{A}_\mathrm{fb}^{-1} &\ge 0, \\
    \mbf{A}_\mathrm{ff}^\trans \mbf{P} \mbf{A}_\mathrm{ff} - \bar{\rho}^4\mbf{A}_\mathrm{bb}^\trans\mbf{P}\mbf{A}_\mathrm{bb} &\ge 0. \label{proof_ass_1} 
\end{align}
Then, adding and subtracting to the left side of \eqref{proof_ass_1} by $\bar{\rho}^2\mbf{P}$ leads to
\begin{align}
   \mbf{A}_\mathrm{ff}^\trans \mbf{P} \mbf{A}_\mathrm{ff} - \bar{\rho}^4\mbf{A}_\mathrm{bb}^{\trans}\mbf{P}\mbf{A}_\mathrm{bb} + \bar{\rho}^2\mbf{P} - \bar{\rho}^2\mbf{P} &\ge 0, \\
   \mbf{A}_\mathrm{ff}^\trans \mbf{P} \mbf{A}_\mathrm{ff} - \bar{\rho}^2\mbf{P} -\bar{\rho}^4(\mbf{A}_\mathrm{bb}^{\trans}\mbf{P}\mbf{A}_\mathrm{bb} - \frac{1}{\bar{\rho}^2}\mbf{P}) &\ge 0.
\end{align}
Let $\mbf{M} = \mbf{A}_\mathrm{ff}^\trans \mbf{P} \mbf{A}_\mathrm{ff} - \bar{\rho}^2\mbf{P}$ and $\mbf{N} = (\mbf{A}_\mathrm{bb}^\trans\mbf{P}\mbf{A}_\mathrm{bb} - \frac{1}{\bar{\rho}^2}\mbf{P})$, then 
\begin{equation}
    \mbf{M} - \bar{\rho}^4\mbf{N} \ge 0, \label{proof_end}
\end{equation}
where \textbf{M} and \textbf{N} are used to represent the left-hand sides of \eqref{forw_lyap} and \eqref{here_43}, respectively. Recall that \eqref{forw_lyap} forces the spectral radius of $\mbf{A}_\mathrm{ff}$ to be bounded above by $\bar{\rho}$, and \eqref{here_43} forces the eigenvalues of $\mbf{A}_\mathrm{bb}$ to be bounded below by $\frac{1}{\bar{\rho}}$. Due to the fact that \eqref{proof_end} is positive semidefinite, either \textbf{M} has to be positive semidefinite or $\bar{\rho}^4\mbf{N}$ has to be negative semidefinite and sufficiently large for \eqref{proof_end} to hold. Therefore, since the contrapositive is assumed and either \textbf{M} does not respect \eqref{forw_lyap} or \textbf{N} does not respect \eqref{here_43}, then, \eqref{lyap_avg} must hold. Note that for this statement to hold, both the forward- and backward-in-time dynamics matrices must be solved with a common Lyapunov variable \textbf{P}.

 To show that the matrix square root operation conserves the bounds constraining the eigenvalues of $\mbf{A}_\mathrm{ff}\mbf{A}_\mathrm{fb}$ above by $\bar{\rho}^2$ in \eqref{paper_unbiased_A}, consider the Schur decomposition~\cite{Deadman2012}
\begin{equation}
    \mbf{A}_\mathrm{ff}\mbf{A}_\mathrm{fb} = \mbf{Q}\mbf{T}\mbf{Q}^{-1},
\end{equation}
where \textbf{Q} is unitary and \textbf{T} is upper triangular with the same eigenvalues as $\mbf{A}_\mathrm{ff}\mbf{A}_\mathrm{fb}$ by similarity transform. Let the square root of \textbf{T} be \textbf{S}, such that
\begin{equation}
    \left(\mbf{A}_\mathrm{ff}\mbf{A}_\mathrm{fb}\right)^\frac{1}{2} = \mbf{Q}\mbf{T}^{\frac{1}{2}}\mbf{Q}^{-1},
\end{equation}
where 
\begin{equation}
    \mbf{S}^2 = \mbf{T}.
\end{equation}
Then, the elements of \textbf{S} are found as~\cite{Deadman2012}
\begin{align}
    S_{ii}^2 &= T_{ii}, \label{diag_eig} \\
    S_{ii}S_{ij} + S_{ij}S_{jj} &= T_{ij} - \sum_{k = i + 1}^{j - 1} S_{ik}S_{kj},
\end{align}
where $S_{ij}$ and $T_{ij}$ represent an element of \textbf{S} and \textbf{T}, respectively.

Note that \textbf{S} is also upper triangular, which means its main diagonal consists of its eigenvalues, which are the square roots of the eigenvalues of \textbf{T} by \eqref{diag_eig}. Since the spectral radius of $\mbf{A}_\mathrm{ff}\mbf{A}_\mathrm{fb}$ is bounded above by $\bar{\rho}^2$, then, the spectral radius of \textbf{S} is bounded above by $\bar{\rho}$.

Therefore, the dynamics matrix with reduced bias, described by \eqref{paper_unbiased_A}, has a spectral radius bounded above by $\bar{\rho}$ as long as the forward- and backward-in-time dynamics matrices respect \eqref{forw_lyap} and \eqref{here_43}, respectively, meaning that the Koopman system with reduced bias is asymptotically stable when $0 < \bar{\rho} < 1$.

\section{Formulation of the optimization problem} \label{optimization_section}
\noindent
The forward-in-time asymptotically stable Koopman system is defined with the Koopman matrix $\mbf{U}_{\mathrm{ff}}$, which can be approximated with BMI constraints as~\cite{Dahdah2022}
\begin{align}
   & \mbf{U}_{\mathrm{ff}} = \argmin_{\mbf{A}_{\mathrm{ff}}, \mbf{B}_{\mathrm{ff}}, \mbf{P}} \hspace{10pt} \left\|\mbs{\Theta}_+ - \begin{bmatrix}
        \mbf{A}_{\mathrm{ff}} & \mbf{B}_{\mathrm{ff}}
    \end{bmatrix}\mbs{\Psi}\right\|^2_\frob \label{vanilla_cost} \\
    & \textrm{s.t.} \hspace{10pt} \mbf{P} > 0, \hspace{10pt}
    \begin{bmatrix}
        \bar{\rho}\mbf{P} & \mbf{A}_{\mathrm{ff}}\mbf{P} \\
        \star &
        \bar{\rho}\mbf{P}
    \end{bmatrix} > 0, \label{C1}
\end{align}
where $0 < \bar{\rho} \le 1$.

In this section, the bilinear Koopman optimization problem described by \eqref{vanilla_cost} and \eqref{C1} is transformed into a convex one to simplify and decrease the solving time. Then, the convex problem is reformulated to solve for both the forward- and backward-in-time Koopman matrices simultaneously.

\subsection{The convex forward-in-time problem} \label{bmi_to_lmi_trick}
\noindent
To make the optimization problem described by \eqref{vanilla_cost} and \eqref{C1} convex, \eqref{C1} must not have bilinear terms. To avoid using the nonlinear term $\mbf{A}_\mathrm{ff}\mbf{P}$,~\cite{Mabrok2023} suggests a transformation that allows \eqref{vanilla_cost} and \eqref{C1} to be rewritten in a linear fashion. Consider the modified cost function with weighting matrix~$\bar{\mbf{W}}$
\begin{equation} \label{mod_cost}
    J(\mbf{A}_{\mathrm{ff}}, \mbf{B}_{\mathrm{ff}}, \mbf{P}) = \left\|\left(\mbs{\Theta}_+ - \begin{bmatrix}
        \mbf{A}_\mathrm{ff} & \mbf{B}_\mathrm{ff}
    \end{bmatrix} \mbs{\Psi}\right)\bar{\mbf{W}}\right\|^2_\frob
\end{equation}
where 
\begin{equation}
    \bar{\mbf{W}} = \mbs{\Psi}^\trans (\mbs{\Psi} \mbs{\Psi}^\trans)^\dagger \bar{\mbf{P}}, \hspace{20pt}
    \bar{\mbf{P}} = \begin{bmatrix} 
        \mbf{P} & \textbf{0} \\ 
        \textbf{0} & \textbf{1}
    \end{bmatrix}. \label{p_const}    
\end{equation}
Substituting $\bar{\mbf{W}}$ in \eqref{mod_cost} gives
\begin{align}
    J(\mbf{A}_{\mathrm{ff}}, \mbf{B}_{\mathrm{ff}}, \mbf{P}) &= \left\|\left(\mbs{\Theta}_+ - \begin{bmatrix}
        \mbf{A}_\mathrm{ff} & \mbf{B}_\mathrm{ff}
    \end{bmatrix} \mbs{\Psi}\right) \mbs{\Psi}^\trans (\mbs{\Psi} \mbs{\Psi}^\trans)^\dagger \bar{\mbf{P}}\right\|^2_\frob \\
    &= \left\|\left(\mbs{\Theta}_+\mbs{\Psi}^\trans (\mbs{\Psi} \mbs{\Psi}^\trans)^\dagger \bar{\mbf{P}} - \begin{bmatrix}
        \mbf{A}_\mathrm{ff} & \mbf{B}_\mathrm{ff}
    \end{bmatrix} \mbs{\Psi}\mbs{\Psi}^\trans (\mbs{\Psi} \mbs{\Psi}^\trans)^\dagger \bar{\mbf{P}}\right)\right\|^2_\frob \\
    &= \left\|\mbf{G}_\mathrm{f} \mbf{H}_\mathrm{f}^\dagger \bar{\mbf{P}} - \begin{bmatrix}
        \mbf{A}_\mathrm{ff}\mbf{P} & \mbf{B}_\mathrm{ff}
    \end{bmatrix}\right\|^2_\frob, \label{here_12}
\end{align}
where $\mbf{G}_\mathrm{f}$ and $\mbf{H}_\mathrm{f}$ are defined in \eqref{forw_edmd}.
The change of variable $\mbf{X}_\mathrm{f} = \mbf{A}_\mathrm{ff}\mbf{P}$ is then used to make \eqref{here_12} linear in $\mbf{X}_\mathrm{f}$, $\mbf{B}_\mathrm{ff}$ and $\bar{\mbf{P}}$, such that 
\begin{equation}
    J(\mbf{A}_{\mathrm{ff}}, \mbf{B}_{\mathrm{ff}}, \mbf{P})
    = \left\|\mbf{G}_\mathrm{f} \mbf{H}_\mathrm{f}^\dagger \bar{\mbf{P}} - \begin{bmatrix}
        \mbf{X}_\mathrm{f} & \mbf{B}_\mathrm{ff}
    \end{bmatrix}\right\|^2_\frob.
\end{equation}
The BMI in \eqref{C1} can then be written as 
\begin{equation}
    \begin{bmatrix}
        \bar{\rho}\mbf{P} & \mbf{X}_\mathrm{f} \\
        \star &
        \bar{\rho}\mbf{P}
    \end{bmatrix} > 0.
\end{equation}
After using the proposed change of variables, the optimization problem described in \eqref{vanilla_cost} and \eqref{C1} becomes
\begin{align}
   & \min \;\;\;\; J(\mbf{X}_\mathrm{f}, \mbf{B}_\mathrm{ff}, \mbf{P}) = \left\|\mbf{G}_\mathrm{f} \mbf{H}_\mathrm{f}^\dagger \bar{\mbf{P}} - \begin{bmatrix}
        \mbf{X}_\mathrm{f} & \mbf{B}_\mathrm{ff}
    \end{bmatrix}\right\|^2_\frob \label{mod_cost_final} \\
    & \textrm{s.t.} \;\;\;\; \mbf{P} > 0, \;\;\;\;
    \begin{bmatrix}
        \bar{\rho}\mbf{P} & \mbf{X}_\mathrm{f} \\
        \star &
        \bar{\rho}\mbf{P}
    \end{bmatrix} > 0, \label{here52}
\end{align}
where $\bar{\mbf{P}}$ is defined in \eqref{p_const}. Then, $\mbf{A}_\mathrm{ff}$ can be solved using $\mbf{A}_\mathrm{ff} = \mbf{X}_\mathrm{f}\mbf{P}^{-1}$, where $\mbf{P}$ is invertible because it is constrained to be positive definite. 

Now that the optimization problem has been rewritten in a convex LMI form, consider the cost function in \eqref{mod_cost_final}. A similar procedure as in~\cite{Dahdah2022}, which introduces a slack variable~\cite[\S 2.15]{Caverly2019} in \eqref{mod_cost_final}, allows the optimization problem described by \eqref{mod_cost_final} and \eqref{here52} to be rewritten as
\begin{align}
& \min \;\;\;\; J(\gamma, \mbf{X}_\mathrm{f}, \mbf{B}_\mathrm{ff}, \mbf{P}, \mbf{Z}) = \gamma \label{slack_cost_no_AS} \\
    & \textrm{s.t.} \;\;\;\; \trace{(\mbf{Z})} < 1, \;\;\;\;
    \mbf{Z} > 0, \;\;\;\;
    \begin{bmatrix}
        \mbf{Z} & \left(\mbf{G}_\mathrm{f} \mbf{H}_\mathrm{f}^\dagger \bar{\mbf{P}} - \begin{bmatrix}
            \mbf{X}_\mathrm{f} & \mbf{B}_\mathrm{ff}
        \end{bmatrix}\right)^\trans \\ \star & \gamma \textbf{1}
    \end{bmatrix} > 0, \\
    & \mbf{P} > 0, \;\;\;\;
    \begin{bmatrix}
        \bar{\rho}\mbf{P} & \mbf{X}_\mathrm{f} \\
        \star &
        \bar{\rho}\mbf{P}
    \end{bmatrix} > 0, \label{slack_cost_no_AS_end}
\end{align}
where it is now formulated in a modular fashion.

\subsection{The combined optimization problem}
\noindent
To solve for both the forward- and backward-in-time Koopman matrices simultaneously, the optimization problem described by \eqref{slack_cost_no_AS}--\eqref{slack_cost_no_AS_end} must be extended to include the backward-in-time dynamics. This extension on the optimization problem is done by adding another slack variable and the appropriate constraints as well as the constraint on the backward-in-time dynamics matrix's eigenvalues in \eqref{back lyap} adapted with the proposed change of variables, such that the problem becomes
\begin{align}
    & \min \hspace{10pt} J(\gamma, \nu, \mbf{X}_\mathrm{f}, \mbf{B}_\mathrm{ff}, \mbf{X}_\mathrm{b}, \mbf{B}_\mathrm{bb}, \mbf{P}, \mbf{Z}, \mbf{V}) = \gamma + \nu \label{final_opt_true} \\
    & \textrm{s.t.} \hspace{10pt} \trace{(\mbf{Z})} < 1, \hspace{10pt}
    \trace{(\mbf{V})} < 1, \hspace{10pt}
    \mbf{Z} > 0, \hspace{10pt}
    \mbf{V} > 0, \hspace{10pt}
    \mbf{P} > 0, \label{old_constr}\\
    & \hspace{20pt}\begin{bmatrix}
        \mbf{Z} & \left(\mbf{G}_\mathrm{f} \mbf{H}_\mathrm{f}^\dagger \bar{\mbf{P}} - \begin{bmatrix}
            \mbf{X}_\mathrm{f} & \mbf{B}_\mathrm{ff}
        \end{bmatrix}\right)^\trans \\ \star & \gamma \textbf{1}
    \end{bmatrix} > 0, \hspace{10pt}
    \begin{bmatrix}
        \mbf{V} & \left(\mbf{G}_\mathrm{b} \mbf{H}_\mathrm{b}^\dagger \bar{\mbf{P}} - \begin{bmatrix}
            \mbf{X}_\mathrm{b} & \mbf{B}_\mathrm{bb}
        \end{bmatrix}\right)^\trans \\ \star & \nu \textbf{1}
    \end{bmatrix} > 0, \label{here22} \\
    & \hspace{20pt}\begin{bmatrix}
        \bar{\rho}\mbf{P} & \mbf{X}_\mathrm{f} \\
        \star &
        \bar{\rho}\mbf{P}
    \end{bmatrix} > 0, \hspace{10pt}
    \bar{\rho}\mbf{X}_\mathrm{b} + \bar{\rho}\mbf{X}_\mathrm{b}^\trans - 2\mbf{P} > 0, \label{here23}
\end{align}
where $\mbf{X}_\mathrm{b} = \mbf{A}_\mathrm{bb}\mbf{P}$, and $\mbf{G}_\mathrm{b}$ and $\mbf{H}_\mathrm{b}$ are defined in \eqref{back_edmd}. After solving the problem, $\mbf{A}_\mathrm{ff}$ and $\mbf{A}_\mathrm{bb}$ can be recovered using $\mbf{A}_\mathrm{ff} = \mbf{X}_\mathrm{f}\mbf{P}^{-1}$ and $\mbf{A}_\mathrm{bb} = \mbf{X}_\mathrm{b}\mbf{P}^{-1}$. Finally, the Koopman matrix with reduced bias can then be computed using \eqref{paper_unbiased_A} and \eqref{avg_B} as $\tilde{\mbf{U}} = \begin{bmatrix}
    \tilde{\mbf{A}} & \tilde{\mbf{B}}
\end{bmatrix}$. This new formulation allows the merged dynamics problem to be solved with a common Lyapunov variable \textbf{P}. The presence of a common Lyapunov variable is critical because it is used to enforce asymptotic stability on the Koopman model with reduced bias as presented in Section \ref{section_asym}. Note that if the asymptotic stability constraints in \eqref{here23} are not used, the forward- and backward-in-time optimization problems can be solved separately.

In practice, solving \eqref{final_opt_true}--\eqref{here23} is numerically challenging due to the fact that $\bar{\mbf{P}}$ scales the cost function, but is also a design variable. To elucidate further, consider the minimization of the cost function in \eqref{mod_cost_final}
\begin{equation} \label{deriv_3}
   \min \;\;\;\; J(\mbf{A}_\mathrm{ff},  \mbf{B}_\mathrm{ff}, \mbf{P}) =  \left\|\left(\mbf{G}_\mathrm{f} \mbf{H}_\mathrm{f}^\dagger - \begin{bmatrix}
        \mbf{A}_\mathrm{ff} & \mbf{B}_\mathrm{ff}
    \end{bmatrix}\right)\bar{\mbf{P}}\right\|^2_\frob,
\end{equation}
which is used in the optimization problem described by \eqref{final_opt_true}--\eqref{here23}. It is important that the focus of the optimization is making $\mbf{G}_\mathrm{f} \mbf{H}_\mathrm{f}^\dagger - \begin{bmatrix}\mbf{A}_\mathrm{ff} & \mbf{B}_\mathrm{ff}\end{bmatrix}$ small and not $\bar{\mbf{P}}$ small. Because $\mbf{P}$ appears in $\bar{\mbf{P}}$ via \eqref{p_const}, in turn it is important $\mbf{P}$ is not small. Ideally, $\mbf{P}$ from \eqref{p_const} would have eigenvalues with similar magnitudes to the magnitudes of the eigenvalues of $\mbs{\Psi}^\trans (\mbs{\Psi} \mbs{\Psi}^\trans)^\dagger$, the other term within $\bar{\mbf{W}}$. This restriction would ensure that $\bar{\mbf{P}}$ has a similar Euclidean norm to $\mbs{\Psi}^\trans (\mbs{\Psi} \mbs{\Psi}^\trans)^\dagger$. Consequently, the constraint $\mbf{P} > 0$ is modified to
\begin{equation} \label{new_lmi}
    \mbf{P} - \epsilon\mbf{1} > 0,
\end{equation}
where one choice of epsilon is $\epsilon = \left\|\mbs{\Psi}^\trans (\mbs{\Psi} \mbs{\Psi}^\trans)^\dagger\right\|_2$. 
A similar approach is used in~\cite{Wang2018} to improve the numerical conditioning of \textbf{P} by constraining it to be within an appropriate range.

\section{Results and discussion}\label{results_section}
\noindent
To assess the performance of the Koopman operator approximation with reduced bias obtained from forward-backward EDMD with inputs and asymptotic stability constraints (fbEDMD-AS), a simulated dataset from a Duffing oscillator and an experimental dataset from a soft robot arm are used to identify a Koopman system representation from data.

In this section, the performance of fbEDMD-AS is compared with EDMD, EDMD with asymptotic stability constraints (EDMD-AS), and forward-backward EDMD with inputs (fbEDMD). In summary, the results are obtained by first, collecting data from a real experiment or from a simulation. Note that if the original dataset is noisy, the noise must be characterized for the proposed method to perform properly. Second, the sampled datasets are corrupted by adding white noise and then normalized using min-max normalization to scale the data between -1 and 1. Then, by choosing a set of lifting functions, the noisy dataset is lifted. The choice of lifting functions can vary between different experiments, since there exist many different sets of lifting functions that can suit a particular dataset. Following Figure \ref{summary_fig}, a copy of the lifted dataset is reversed in time to obtain the backward snapshot matrix. The next step involves solving for the forward- and backward-in-time Koopman matrices using the different optimization methods.  A Koopman matrix identified with EDMD is computed using \eqref{forw_edmd1} through \texttt{pykoop}~\cite{Pykoop2021}, an open-source Koopman operator identification library. Similarly, fbEDMD uses \texttt{pykoop} to separately solve the two EDMD problems found in \eqref{forw_edmd1} and \eqref{back_edmd1}, then computes the Koopman matrix with reduced bias using \eqref{paper_unbiased_A} and \eqref{avg_B}. Lastly, EDMD-AS identifies the Koopman matrix using the optimization problem developed in \eqref{slack_cost_no_AS}--\eqref{slack_cost_no_AS_end}, while fbEDMD-AS uses the optimization problem described by \eqref{final_opt_true}--\eqref{here23}. Both EDMD-AS and FBEDMD-AS use the asymptotic stability constraints with the bound $\bar{\rho} = 0.9999$. The bound on the upper eigenvalue can be changed to obtain eigenvalues that are closer or further from the unit circle. Note that the methods posed as semidefinite programming problems are formulated using the PICOS Python library \cite{Picos} and solved with the MOSEK solver \cite{Mosek}.  Finally, once the forward- and backward-in-time Koopman matrices are computed, the unbiased Koopman matrix is obtained using the mathematical expressions displayed in Figure \ref{summary_fig} (\emph{c}), which are derived in Section \ref{fbEDMD}.

To evaluate the performance of the different methods, the resulting Koopman matrices are compared based on their trajectory errors, spectral radii, and accuracy to the noiseless solution. Note that when the spectral radius of the Koopman matrix is referred to, it is the dynamics matrix that is discussed. The code used to generate all the figures presented in this section can be found in the GitHub repository at \url{https://github.com/decargroup/forward_backward_koopman.git}.

\subsection{Duffing Oscillator Simulation}
\noindent
The Duffing oscillator results highlight the effect of the bias reduction step in the trajectory of the system. The continuous-time Duffing oscillator dynamics are \cite{Kovacic2011}
\begin{equation} \label{Duff}
    m\ddot{x}(t)+ c\dot{x}(t) + k_1x(t) + k_2x(t)^3 = f(t),
\end{equation}
where the mass $m = 0.1$ kg, damping coefficient $c = 0.01$ Ns/m, linear spring constant $k_1 = 0.1$ N/m and nonlinear spring constant $k_2 = 0.001$ $\text{N/m}^3$. Note that the simulated dataset of the Duffing oscillator is obtained by discretizing \eqref{Duff} with a time interval of 0.01 s using a forward Euler method \cite{LortieThesis}.

To simulate realistic noise levels, let the measurement noise $\mbf{v}_k$ found in \eqref{output_eq} be simulated as normally distributed noise, where $\mbf{v}_k \sim \mathcal{N}(\mbf{0}, (\sqrt{2}/10)^2 \mbf{1})$. The chosen lifting functions for this simulation consist of all the possible second-order monomial combinations of the states followed by ten thin plate radial basis functions defined as 
\begin{equation} \label{lift_func}
    \psi^\mathrm{rbf}_i(\mbf{x}) = r_i^2\ln\left(r_i\right), \;\;\;\; i = 1, \dots, 10,
\end{equation}
where 
\begin{equation}
    r_i = \alpha\left\|\mbs{\psi}^\mathrm{poly}(\mbf{x}) - \mbf{c}_i\right\| + \delta,
\end{equation}
$\mbs{\psi}^\mathrm{poly}(\mbf{x})$ is a vector made from the possible second-order monomial combinations, $\mbf{c}_i$ is a Latin hypercube sampled center, $\alpha = 0.1$ is a shape parameter used for scaling and $\delta = 0.001$ is an offset to ensure the computation of a defined number. In total, including the displacement and velocity of the oscillator, which act as this system's states, 15 lifting functions are used to identify the Koopman matrix. This dataset consists of 20 training episodes and 2 test episodes.

\begin{figure}
    \centering
    \includegraphics[scale=0.99]{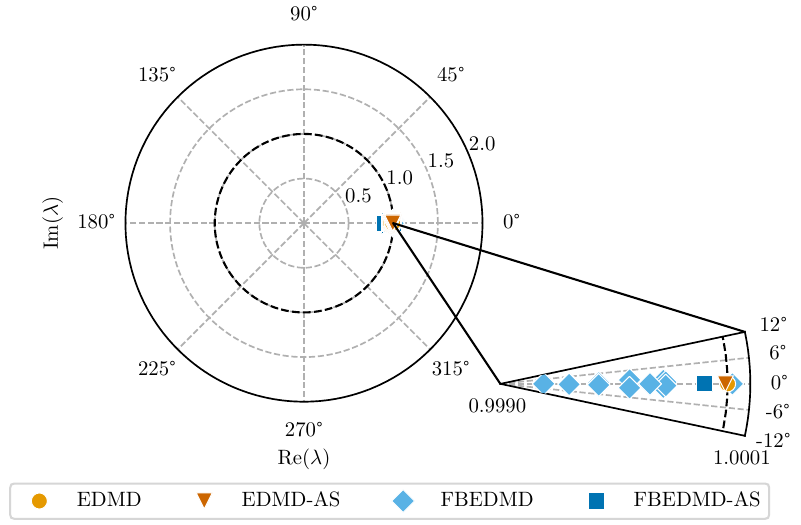}
    \caption{Eigenvalues of the Koopman dynamics matrix identified with the Duffing oscillator dataset. The dynamics matrix identified with fbEDMD has eigenvalues outside the unit circle and therefore produces an unstable model, while the Koopman systems identified with EDMD, EDMD-AS, and fbEDMD-AS are asymptotically stable, since their respective dynamics matrix has its largest eigenvalue within the unit circle.}
    \label{eigen_plot}
\end{figure}

To show the impact of imposing a stability constraint on the problem, the eigenvalues of Koopman matrices solved with EDMD, EDMD-AS, fbEDMD, and fbEDMD-AS are displayed in Figure \ref{eigen_plot}. As opposed to fbEDMD-AS, fbEDMD identifies a Koopman matrix with an eigenvalue outside the unit circle in Figure \ref{eigen_plot}, implying that fbEDMD identified an unstable Koopman system. Although the Duffing oscillator described in \eqref{Duff} is an asymptotically stable system, the choice of lifting functions leads to numerical challenges, which results in an unstable Koopman system. Note that although the eigenvalues of the Koopman matrix identified with EDMD are all within the unit circle, it is possible that some of them be outside the unit circle for a different dataset because no asymptotic stability constraint is enforced.

\begin{figure}
     \centering
         \includegraphics[scale=0.99]{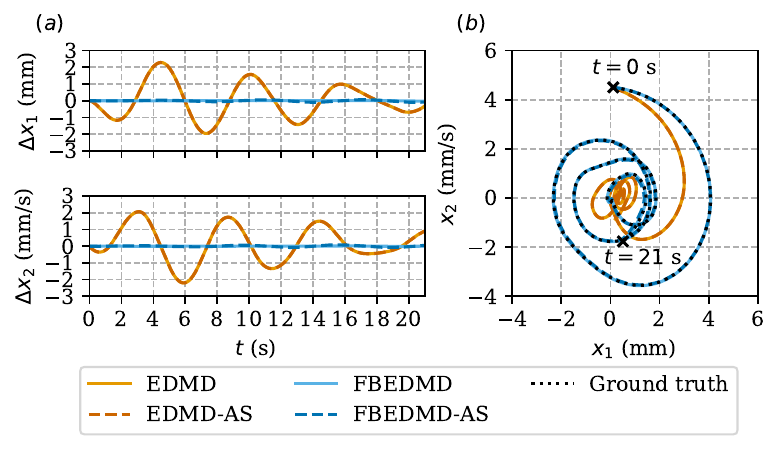}
        \caption{Prediction error plot $(a)$ and multi-step trajectory $(b)$ of Koopman models approximated with the simulated Duffing oscillator dataset. The states $x_1$ and $x_2$ are chosen to be the displacement and velocity of the Duffing oscillator system, respectively. The states are recovered and re-lifted between prediction time steps.
        %
        %
        Bias in the eigenvalues of dynamics matrices identified with EDMD and EDMD-AS leads to overestimated decay rates. The trajectories obtained with the Koopman systems identified with fbEDMD and fbEDMD-AS track better the true trajectory, since most of the bias in the models is removed.}
        \label{trajectory}
\end{figure}

Forward-backward EDMD methods solve for Koopman matrices with reduced bias, which is critical when studying the physical properties of a system, such as the growth and decay rates of its dynamic modes. When comparing the trajectories obtained with methods using fbEDMD and EDMD, Figure \ref{trajectory} shows that the Koopman matrices computed with EDMD and EDMD-AS have higher decay rates than with fbEDMD and fbEDMD-AS. Trajectories obtained with fbEDMD and fbEDMD-AS follow the true trajectory much more accurately, which indicates that the bias is reduced. Furthermore, the prediction errors obtained with fbEDMD methods in Figure \ref{trajectory} are much lower than with EDMD methods, which is another indication that the bias is reduced. Note that although the Koopman system identified with fbEDMD has a smaller prediction error than with fbEDMD-AS, it is unstable and as time grows, the simulated trajectory will likely diverge.

\subsection{Soft Robot Experiment}
\noindent
 Datasets of the soft robot shown in~\cite{Bruder2019, Bruder2020} are used in this section. This soft robot arm has a laser pointer mounted as an end effector, which is used to aim at a planar surface to obtain the 2D coordinates of the resulting point. These coordinates are then used as the system's states. To obtain a desired trajectory, the soft robot is controlled by three pneumatic actuators, which serve as the system's inputs. Although any experimental data is naturally noisy, 
 simulated noise is added to the measurement noise $\mbf{v}_k$ found in \eqref{output_eq} to highlight the differences between the methods. The lifting functions used for this section also consist of all the possible second-order monomial combinations of the states followed by ten radial basis functions generated using \eqref{lift_func} with shape parameter $\alpha = 0.5$. Therefore, the total quantity of lifting functions used for this dataset is 15, including the two states $x_1$ and $x_2$, which are the laser pointer's coordinates in the $x$-axis and $y$-axis, respectively. For this dataset, there are 13 training episodes and 4 test episodes~\cite{Bruder2019, Bruder2020}. 

In this section, signal-to-noise ratios (SNR) of the soft robot dataset, computed as \cite{Oppenheim1999}
\begin{equation}
    \text{SNR} = 10\log\left(\frac{\sigma_x^2}{\sigma_n^2}\right),
\end{equation}
where $\sigma_x$ is the standard deviation of the unlifted signal and $\sigma_n$ is the standard deviation of the added simulated noise, are used to compare the ability of each method to compute a Koopman matrix with reduced bias at certain noise levels. When computing SNR, it is assumed that the original dataset is noise-free. Note that the SNR level associated with the added simulated noise $\mbf{w}_k$ defined as $\mbf{w}_k \sim \mathcal{N}(\mbf{0}, (\sqrt{2}/10)^2 \mbf{1}))$ is approximately 28 dB. Additionally, the relative error between an approximated Koopman operator and the true Koopman operator approximation is used as an accuracy metric. Here, the true Koopman operator approximation is defined as the Koopman matrix obtained with no added simulated noise using the corresponding method.
\begin{figure}
    \centering
    \includegraphics[scale=0.98]{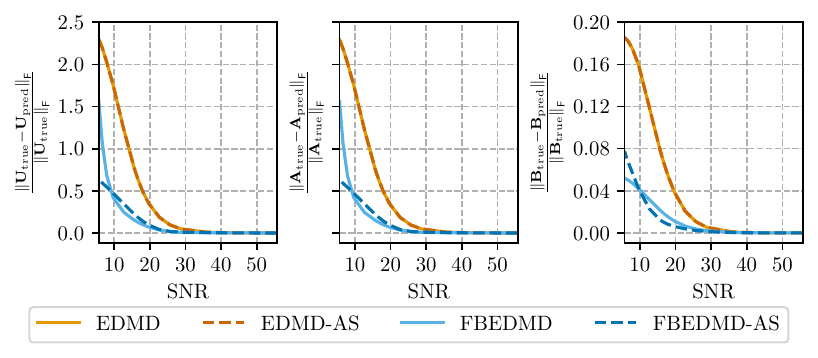}
    \caption{The relative Frobenius norm error of the approximated Koopman, dynamics, and input matrices at different signal-to-noise ratios (SNRs). A Koopman model approximated with FBEDMD methods is more accurate than with EDMD methods under 35 dB. Although FBEDMD-AS uses additional constraints compared to FBEDMD, it approximates a Koopman model with a similar accuracy to FBEDMD.}
    \label{frob_norm_error}
\end{figure}

\begin{figure}
    \centering
    \includegraphics[scale=0.98]{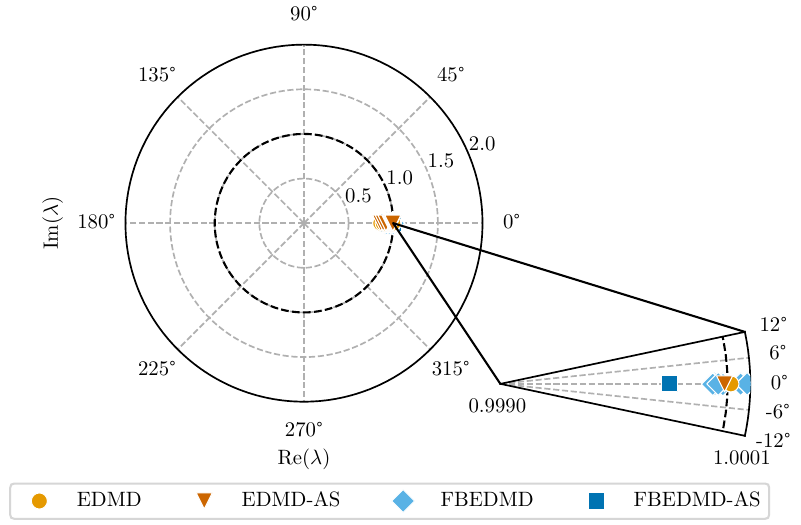}
    \caption{Eigenvalues of the Koopman dynamics matrix identified with the soft robot dataset at an SNR of 28 dB. Both dynamics matrices identified with EDMD and fbEDMD have eigenvalues outside the unit circle and therefore produce unstable models. Koopman systems identified with EDMD-AS and fbEDMD-AS are asymptotically stable, since their respective dynamics matrix has its largest eigenvalues bounded within the unit circle.}
    \label{soft_eigen_plot}
\end{figure}

The results shown in Figure \ref{frob_norm_error} highlight that fbEDMD methods solve for a Koopman matrix that is much closer, in a Frobenius norm sense, to the true Koopman matrix solution than EDMD methods when under a SNR of approximately 35 dB. At higher SNRs, when less noise is added, Figure \ref{frob_norm_error} shows that all methods solve for a Koopman matrix with high accuracy. This result is expected, since the true Koopman matrix is computed using data that has not been artificially corrupted by noise. Although fbEDMD-AS has additional constraints to ensure asymptotic stability, it follows a similar accuracy to fbEDMD. Forward-backward EDMD shows a lower error than fbEDMD-AS between approximately 10 dB and 25 dB, since it is not subject to additional constraints. However, the bias reduction process becomes harder to execute when systems become more unstable, because the relations between the forward- and backward-in-time dynamics found in \eqref{iden_1}--\eqref{iden_4} start to hold only approximately. When SNR is small, below approximately 10 dB, fbEDMD identifies unstable Koopman systems with eigenvalues that are much greater than one, leading to a poor cancellation of the bias and a higher relative error than with fbEDMD-AS. The ability of fbEDMD-AS to identify an asymptotically stable Koopman system with reduced bias makes it more adequate than fbEDMD, which identified an unstable system as shown in Figure \ref{soft_eigen_plot}. Additionally, Figure \ref{frob_norm_error} shows that the reduction in bias with fbEDMD methods is not only observed in the dynamics matrix, but also in the input matrix, which is an expected result.

\begin{figure}
     \centering
         \centering
         \includegraphics[scale=0.98]{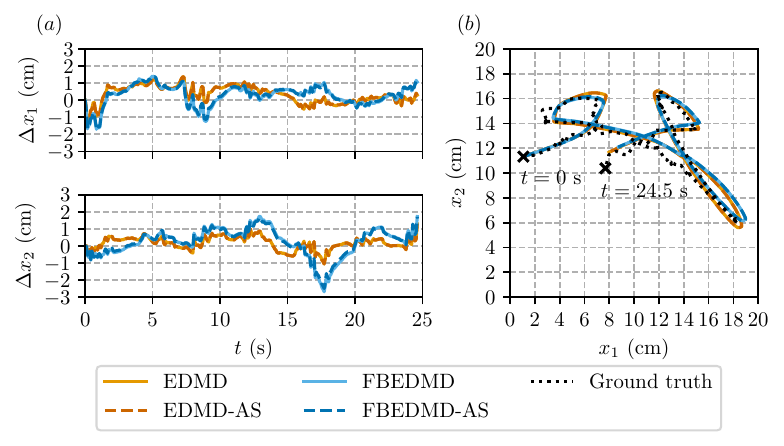}
        \caption{Prediction error plot $(a)$ and multi-step trajectory $(b)$ of the third test episode for Koopman models approximated with the soft robot dataset at an SNR of 28 dB. The states of the soft robot system, $x_1$ and $x_2$, are chosen to be the displacements in the $x$-axis and $y$-axis direction. States are recovered and re-lifted between prediction timesteps.
         The Koopman models identified with EDMD methods and fbEDMD methods have similar prediction errors.}
        \label{soft_trajectory}
\end{figure}

\begin{figure}
     \centering
         \includegraphics[scale=0.97]{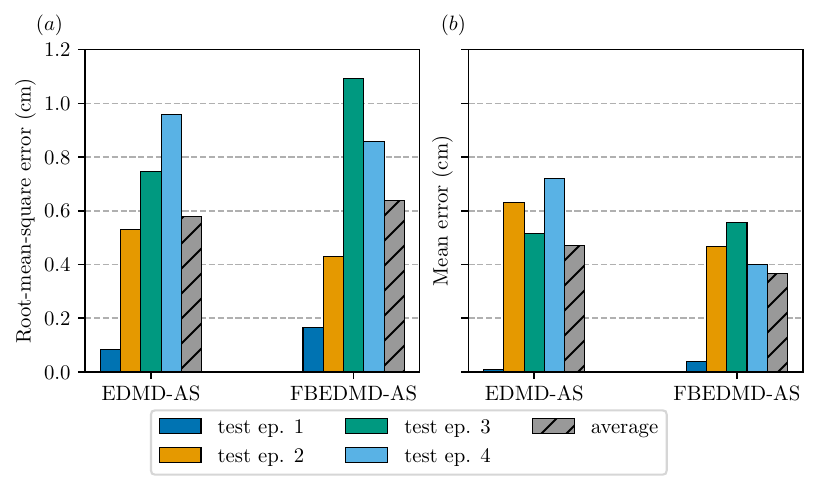}
        \caption{The root-mean-square (RMS) $(a)$ and mean $(b)$ multi-step prediction errors of four test episodes for Koopman models approximated with the soft robot dataset at an SNR of 28 dB. The RMS and mean errors of Koopman systems identified with EDMD and fbEDMD are not shown because they are unstable. Koopman systems identified with fbEDMD-AS have a higher averaged RMS prediction error, but a lower averaged mean error than with EDMD-AS.}
        \label{error_bars}
\end{figure}

The trajectory of the third test episode in Figure \ref{soft_trajectory} does not show that EDMD methods have a significant bias compared to fbEDMD methods, as opposed to Figure \ref{trajectory}. Although the unstable Koopman systems identified with EDMD and fbEDMD have a similar trajectory to EDMD-AS and fbEDMD-AS, respectively, their unstable nature could lead to unbounded prediction errors with different datasets. To better examine the trajectory, the root-mean-square (RMS) error is used to compare the dispersion, and the mean error is used to analyze the bias in the prediction results. When comparing those errors, EDMD and fbEDMD methods are disregarded, as Figure \ref{soft_eigen_plot} shows that they both generate an unstable model. For this dataset, Figure \ref{error_bars} shows that the Koopman matrix computed with fbEDMD-AS does not exhibit a more accurate trajectory than EDMD-AS for each test episode. 
The averaged RMS error in Figure \ref{error_bars} obtained with fbEDMD-AS is higher than with EDMD-AS, while the averaged mean error is smaller for fbEDMD-AS than with EDMD-AS. The results highlight that on average, although trajectory predictions obtained with fbEDMD-AS have more dispersed errors than with EDMD-AS, they have a lower bias than with EDMD-AS, which is the intended result.

\begin{figure}
     \centering
         \includegraphics[scale=0.99]{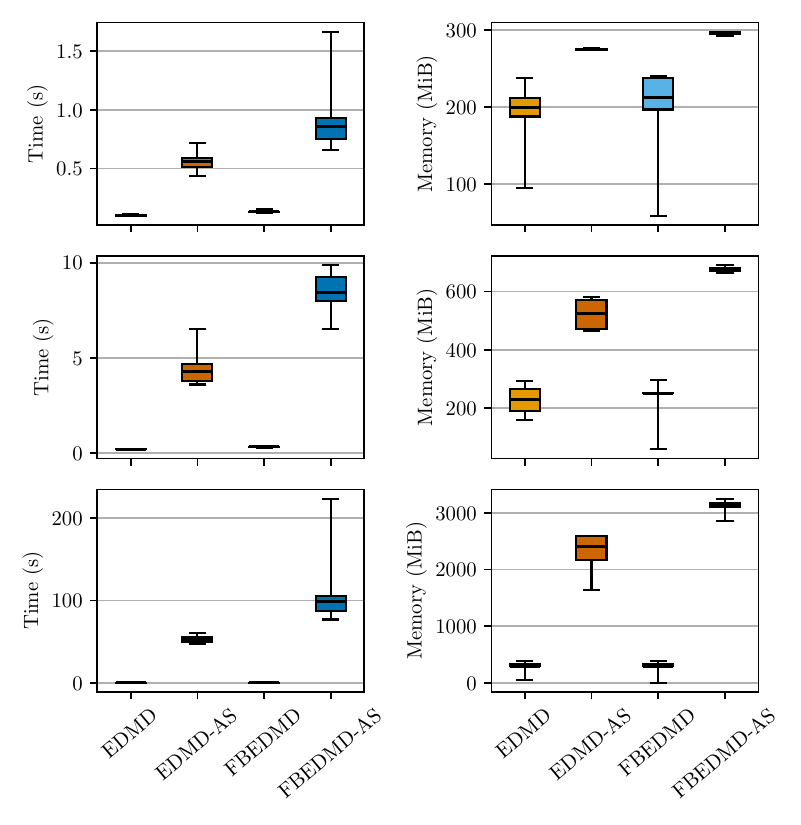}
        \caption{The computation time (left column) and maximum memory usage (right column) when solving for the Koopman matrix with 15 lifting functions (first row), 30 (second row), and 55 lifting functions (third row). Both EDMD and FBEDMD solve for a Koopman matrix using linear least-squares, while EDMD-AS and FBEDMD-AS are formulated semidefinite. The mean (black line) and boxplots of each subplot boxplots are computed using 30 runs for each method. The lower and higher whiskers of each box represent the minimum and maximum values, while the box represents the middle 50$\%$ of the data. In general, EDMD-AS and FBEDMD-AS have higher computational times and maximum memory usage over all tested quantity of lifting functions than EDMD and FBEDMD. Additionally, FBEDMD methods require higher computational times than EDMD methods.}
        \label{cost_mem}
\end{figure}

The proposed method, FBEDMD-AS, combines the forward- and backward-in-time regression problems, which is expected to be more computationally demanding than the forward-in-time method, EDMD-AS. As highlighted in Figure \ref{cost_mem}, not only does FBEDMD-AS require more time to compute a solution, but it also consumes more memory than EDMD-AS. When computing the Koopman matrix with the optimization method EDMD-AS, the solver iterates over 562 variables and 5 constraints as opposed to FBEDMD-AS, which uses 1004 variables and 9 constraints. Additionally, when comparing linear least-square problems, such as EDMD or FBEDMD, to convex optimization problems with LMI constraints, such as EDMD-AS and FBEDMD-AS, it is expected to observe in Figure \ref{cost_mem} a significant change in computing time and memory.  Therefore, for the considered case of 15 lifting functions, FBEDMD-AS requires more computing time and memory to compute a solution, but the difference is not practically important. 

Another component that can impact the computational complexity of the problem is high-dimensional data. To present the performances of all four methods when used to identify Koopman systems with higher dimensional data, Figure \ref{cost_mem} also highlights the outcome of using more radial basis functions as lifting functions to increase the dimensionality of the data. In particular, the performances of all methods when computing a Koopman matrix with 15 and 40 more radial basis functions than the baseline are presented in Figure \ref{cost_mem}. The proposed method increases significantly more in computing time than with EDMD and FBEDMD when more lifting functions are used. However, the state-of-the-art method, EDMD-AS, also increases significantly in time. Therefore, it is expected for both EDMD-AS and FBEDMD-AS methods to be more computationally demanding than EDMD and FBEDMD when increasing the dimensionality of the data because they are posed as semidefinite problems. Accordingly, depending on the application, the proposed method is not always the best solution if computing time is important, since it increases significantly with more lifting functions. 

The assumption that the added noise is independent and identically distributed with zero mean and finite variance even after the transformation through nonlinear lifting functions is required to obtain the results displayed in this section. Although this assumption may not hold in practice, the results presented in this section indicate that good prediction results can still be obtained on the datasets considered.

\section{Conclusion}\label{conclusion_section}
\noindent
Identifying an accurate Koopman model from noisy input-output data is a difficult task to accomplish because of the introduced bias, which can have a major impact on the growth and decay rates. The proposed method, fbEDMD-AS, reduces the bias from both the dynamics and input matrices of the model, while enforcing asymptotic stability on the Koopman system. The novel contributions of this paper are the introduction of inputs to fbEDMD and the application of asymptotic stability constraints to the model with reduced bias. With datasets from a simulated Duffing oscillator and experimental soft robot arm, fbEDMD-AS is shown to outperform the state-of-the-art EDMD-based methods when attempting to synthesize an accurate Koopman model from noisy data. 

The fbEDMD-AS approach does not guarantee better trajectory prediction for all datasets, since a system that requires an extensive amount of lifting functions for proper identification develops numerical issues causing the relations from \eqref{iden_1}--\eqref{iden_4} to be inconsistent with each other. Future work will be oriented towards overcoming this limitation by using a similar approach as explored in~\cite{Dahdah2022}, where $\mathcal{H}_\infty$ and potentially $\mathcal{H}_2$ norm regularizers are applied to the Koopman operator approximation to enforce asymptotic stability on the Koopman system and improve the conditioning of the Koopman matrix. Furthermore, the proposed method can only be used when the lifting functions used to identify the input matrix are a function of only the inputs. In future work, new solutions that allow the use of states and inputs in the lifting functions used with the input matrix will be derived to compute the dynamics and input matrices with reduced bias. 
\vskip6pt

\section*{Appendix} \label{Appendix}\noindent
To ensure that the forward-in-time dynamics of a system are asymptotically stable using the backward-in-time dynamics, \eqref{here_43} must hold. This appendix shows that \eqref{here_43} is equivalent to bounding the eigenvalues of $\mbf{A}_\mathrm{bb}$ below by $\frac{1}{\bar{\rho}}$ when $\mbf{P} > 0$. Let the eigenvalue equation of $\mbf{A}_\mathrm{bb}$ and its conjugate transpose form be
\begin{equation}
    \mbf{w}^\herm\mbf{A}_\mathrm{bb} = \lambda\mbf{w}^\herm, \label{eigen_val1}
\end{equation}
and
\begin{equation}
    \mbf{A}_\mathrm{bb}^\trans\mbf{w} = \lambda^\ast\mbf{w}, \label{eigen_val2}
\end{equation}
respectively, where $(\cdot)^\herm$ denotes the Hermitian, $(\cdot)^\ast$ denotes the complex conjugate, $\mbf{w}$ is a left eigenvector of $\mbf{A}_\mathrm{bb}$ and $\lambda$ is an eigenvalue of $\mbf{A}_\mathrm{bb}$. Note that for this paper, $\mbf{A}_\mathrm{bb}$ is a real matrix, such that $\mbf{A}^\ast_\mathrm{bb} = \mbf{A}_\mathrm{bb}$. Then, using a similar approach as in \cite[\S 6]{Williams2007} to derive the bounds on the eigenvalues of $\mbf{A}_\mathrm{ff}$ in continuous-time, let \eqref{here_43} be premultiplied and postmultiplied by $\mbf{w}^\herm$ and $\mbf{w}$, respectively, yielding 
\begin{equation}
    \mbf{w}^\herm\mbf{A}_\mathrm{bb}\mbf{P}\mbf{A}_\mathrm{bb}^\trans\mbf{w} - \frac{1}{\bar{\rho}^2}\mbf{w}^\herm\mbf{P}\mbf{w} > 0. \label{door_1}
\end{equation}
Substituting both \eqref{eigen_val1} and \eqref{eigen_val2} into \eqref{door_1} results in
\begin{align}
    \lambda\mbf{w}^\herm\mbf{P}\mbf{w}\lambda^\ast - \frac{1}{\bar{\rho}^2}\mbf{w}^\herm\mbf{P}\mbf{w} > 0, \\
    \left(\lambda\lambda^\ast - \frac{1}{\bar{\rho}^2}\right)\mbf{w}^\herm\mbf{P}\mbf{w} > 0, \\
    \left(\left|\lambda\right|^2 - \frac{1}{\bar{\rho}^2}\right)\mbf{w}^\herm\mbf{P}\mbf{w} > 0.
    \label{door_2}
\end{align}
For \eqref{door_2} to hold, $\left|\lambda\right| > \frac{1}{\bar{\rho}}$ must be true, since $\mbf{P} > 0$. Therefore, it is shown that \eqref{here_43} is equivalent to bounding the eigenvalues of $\mbf{A}_\mathrm{bb}$ below by $\frac{1}{\bar{\rho}}$.

\section{Funding} This work was supported by CIFAR, the Natural Sciences and Engineering Research Council of Canada (NSERC) discovery grants program, the Fonds de recherche du Québec – Nature et technologies (FRQNT), the Toyota Research Institute, the National Science Foundation Career Award (grant no.
1751093) and the Office of Naval Research (grant no. N00014-18-1-2575).

\bibliographystyle{unsrt}
\bibliography{fbedmd}

\end{document}